\documentclass[12pt,preprint]{aastex}
\def \la{\mathrel{\mathchoice   {\vcenter{\offinterlineskip\halign{\hfil
$\displaystyle##$\hfil\cr<\cr\sim\cr}}}
{\vcenter{\offinterlineskip\halign{\hfil$\textstyle##$\hfil\cr
<\cr\sim\cr}}}
{\vcenter{\offinterlineskip\halign{\hfil$\scriptstyle##$\hfil\cr
<\cr\sim\cr}}}
{\vcenter{\offinterlineskip\halign{\hfil$\scriptscriptstyle##$\hfil\cr
<\cr\sim\cr}}}}}
\def \ga{\mathrel{\mathchoice {\vcenter{\offinterlineskip\halign{\hfil
$\displaystyle##$\hfil\cr>\cr\sim\cr}}}
{\vcenter{\offinterlineskip\halign{\hfil$\textstyle##$\hfil\cr
>\cr\sim\cr}}}
{\vcenter{\offinterlineskip\halign{\hfil$\scriptstyle##$\hfil\cr
>\cr\sim\cr}}}
{\vcenter{\offinterlineskip\halign{\hfil$\scriptscriptstyle##$\hfil\cr
>\cr\sim\cr}}}}}

\def\ffam  {\hbox{$\,.\!\!^{\prime}$}}

\begin{document}

\shortauthors{Wang et al.} \shorttitle{Abundances and isotope ratios
in the Large Magellanic Cloud}

\title{Abundances and Isotope Ratios in the Magellanic Clouds:
       The Star Forming Environment of N\,113\altaffilmark{6}}

%\subtitle{}

\author{M.~Wang\altaffilmark{1,2},
        Y.-N. Chin\altaffilmark{3},
        C.~Henkel\altaffilmark{2},
        J. B.~Whiteoak\altaffilmark{4},
        M. Cunningham\altaffilmark{5}}

%\offprints{M.~Wang}

\altaffiltext{1}{Purple Mountain Observatory, 2 West Beijing Road,
  210008 Nanjing, China}\email{mwang@pmo.ac.cn}
\altaffiltext{2}{Max-Planck-Institut f{\"u}r Radioastronomie,
           Auf dem H{\"u}gel 69, D-53121 Bonn, Germany}
\altaffiltext{3}{Department of Physics, Tamkang University,
           251-37 Tamsui, Taipei County, Taiwan}
\altaffiltext{4}{Australia Telescope National Facility, CSIRO
Radiophysics Labs. P.O. Box 76, Epping, NSW 2121, Australia; vis.
1966 Coomba Rd, Coomba Park, NSW 2428, Australia}
\altaffiltext{5}{School of Physics, University of New South Wales
(UNSW), 2052 Sydney, Australia}

      \altaffiltext{6}{Based on observations with the Swedish/ESO Submillimeter Telescope (SEST)
      at the European Southern Observatory (ESO, La Silla, Chile) and the Atacama Pathfinder EXperiment (APEX, Chajnantor,
      Chile) of the Max-Planck-Institut f{\"u}r Radioastronomie (MPIfR), ESO, and Onsala Space Observatory
      (OSO)}
      %\thanks{Figures~A.1 and B.1 -- B.5 and Table ~B.1 are available in electronic form at http://www.edpsciences.org}}

%\titlerunning{Abundances and isotope ratios in the Large Magellanic Cloud}

%\authorrunning{Wang et al.}
%\date{Received date ; accepted date}

\begin{abstract}

With the goal of deriving the physical and chemical conditions
of star forming regions in the Large Magellanic Cloud (LMC), a
spectral line survey of the prominent star forming region N113 is
presented. The observations cover parts of the frequency range
from 85\,GHz to 357\,GHz and include 63 molecular transitions from a
total of 16 species, among them spectra of rare isotopologues. Maps
of selected molecular lines as well as the 1.2\,mm continuum
distribution are also presented. Molecular abundances in the core of
the complex are consistent with a photon dominated region (PDR) in a
nitrogen deficient environment. While CO shows optical depths of
order $\tau$$\sim$10, $^{13}$CO is optically thin. The most
prominent lines of CS, HCN, and HCO$^+$ show signs of weak saturation
($\tau$$\sim$0.5). Densities range from 5$\times$10$^3$\,cm$^{-3}$
for CO to almost 10$^6$ for CS, HCN, and a few other species,
indicating that only the densest regions provide sufficient
shielding even for some of the most common species. An ortho- to
para-H$_2$CO ratio of $\sim$3 hints at H$_2$CO formation in 
a warm ($\ga$40\,K) environment. Isotope ratios are $^{12}$C/$^{13}$C
$\sim$ 49$\pm$5, $^{16}$O/$^{18}$O $\sim$ 2000$\pm$250,
$^{18}$O/$^{17}$O $\sim$ 1.7$\pm$0.2 and $^{32}$S/$^{34}$S $\sim$
15. Agreement with data from other star forming clouds shows
that the gas is well mixed in the LMC. The isotope ratios do not 
only differ from those seen in the Galaxy. They also do not form 
a continuation of the trends observed with decreasing metallicity from
the inner to the outer Galaxy. This implies that the outer Galaxy,
even though showing an intermediate metallicity, is not providing a
transition zone between the inner Galaxy and the metal poor
environment of the Magellanic Clouds. A part of this discrepancy is
likely caused by differences in the age of the stellar populations
in the outer Galaxy and the LMC. While, however, this scenario
readily explains measured carbon and oxygen isotope ratios, nitrogen
and sulfur still lack a self-consistent interpretation.

\end{abstract}


%\maketitle

\section{Introduction}

The Magellanic Clouds are two southern irregular galaxies that
provide unique opportunities to study astrophysical processes (e.g.,
Westerlund 1990). Noteworthy are their extremely small distances
($\sim$50 and 60\,kpc), low heavy element contents, low dust-to-gas
mass ratios, high [O/C] and [O/N] elemental abundance ratios, high
atomic-to-molecular hydrogen (H/H$_2$) ratios, and an intense
ultraviolet (UV) and far-ultraviolet (FUV) radiation field. The
Magellanic Clouds, which are much smaller than the Milky Way, are
characterized by a lower degree of nuclear processing than the
Galaxy. Right now, however, they are undergoing an episode of
vigorous star formation.

In the past couple of decades there have been an enormous number of
studies of the Magellanic Clouds (see, e.g., the IAU Symposia No.
108, 148 190, and 256). The first systematic molecular surveys,
covering the central parts of the Magellanic Clouds in the CO
$J$=1--0 line were those of Cohen et al. (1988) and Rubio et al.
(1991) with angular resolutions of 8\ffam8. More detailed CO
surveys were carried out with the NANTEN telescope at 2\ffam6
resolution (e.g., Yamaguchi et al. 2001; Mizuno et al. 2006). CO
data with even higher resolution, obtained in the $J$=1--0 and 2--1
lines with 50$''$ and 25$''$ beamwidth, were taken with the SEST
(Swedish-ESO Submillimeter Telescope). Several articles were published
as part of the SEST Key-Program on CO in the Magellanic Clouds (e.g.,
Israel et al. 2003).

While these studies provide an overall view of the well shielded
molecular medium in the Magellanic Clouds, multiline studies
investigating the physical and chemical properties of the gas only
exist for a very small number of targets in the vicinity of
prominent HII regions. Following the pioneering study of Johansson
et al. (1994) on N\,159 in the LMC, Chin et al. (1997, 1998) and
Heikkil{\"a}, et al. (1998, 1999) published molecular multi-line
studies on star forming regions in the Magellanic Clouds. First
detections of deuterated molecules outside the Galaxy were
reported by Chin et al. (1996b) and Heikkil{\"a}, et al. (1997).

N\,113, located in the central part of the LMC more than 2$^{\circ}$
west of 30\,Dor, is hosting the most intense H$_2$O maser of the
Magellanic Clouds (Whiteoak \& Gardner 1986; Lazendic et al. 2002;
Oliveira et al. 2006). OH maser emission was also observed
(Brooks \& Whiteoak 1997). An IRAS (InfraRed Astronomy Satellite)
point source, IRAS~0513--694, is associated with N\,113. While
Chin et al. (1996b, 1997) and Wong et al. (2006) collected
single-dish and interferometric data from some $\lambda$3\,mm
transitions, a dedicated molecular multiline study of this star
forming region covering the $\lambda$1--3\,mm band is still
missing. To improve our understanding of the interstellar medium
(ISM) associated with massive star formation in an environment
with low metallicity and to obtain a comprehensive view onto
one of the most remarkable star forming regions of the LMC, we
mapped the 1.2\,mm continuum, obtained a spectral survey of the
central region, and also mapped the cloud in several molecular
lines.

\section{Observations}

\subsection{1.2\,mm continuum observations}

In October 2002 and August 2003 the 15-m SEST Imaging Bolometer
Array SIMBA was used to image the distribution of continuum emission
at 250\,GHz ($\lambda$1.2\,mm) associated with N113. The 37-channel
instrument was configured to map an area with dimensions of
400\arcsec\ in azimuth and 392\arcsec\ in elevation by means of 51
successive azimuth scans, at 80\arcsec/sec, with a separation of
8\arcsec\ in elevation. Including overhead, a completed map took
about six minutes. A total of 32 maps were obtained, 20 during the
first observing period and 12 during the second.

At 250 GHz, the SEST beam had a full width to half maximum (FWHM) of
$\sim$24\arcsec. Accurate pointing and focus of the antenna were
maintained with periodic observations of point-source calibrators
(mainly OA\,129).  Periodic ‘skydip’ observations provided
estimates of the sky opacity. The derived zenith optical depths for
both observing periods averaged about 0.14. Flux density calibration
was provided by observations of Uranus, and were based on an adopted
brightness temperature of 93\,K for the planet.

The observations were processed using MOPSI, a software program
developed and upgraded by R. Zylka (IRAM/Grenoble). Images were
produced for each set of scans, and these were averaged to produce
an image gridded in right ascension and declination. To reduce
systematic baseline effects, the area containing the main source
emission was defined by a polygon boundary, and this area was
excluded from baselining corrections in a second averaging process.
The algorithm PLANET enabled a flux density/beam scale to be derived
from the Uranus observations.

\subsection{Spectroscopic measurements}

\subsubsection{SEST 15-m observations}

With the SEST, observations were carried out in January and
September 1995, January and March 1996, January, March, July, and
September 1997, July 1998, and July 1999. For the frequency ranges,
single-sideband system temperatures on a main beam brightness
temperature scale ($T_{\rm mb}$), beam widths and linear resolutions,
see Table~\ref{tel.param}. 3 and 2\,mm or 3 and 1.3\,mm SIS receivers
were employed simultaneously. For the SIS receiver used at
330--357\,GHz, see also Mauersberger et al. (1996b).

The backend was an acousto-optical spectrometer (AOS) which was
split into 2$\times$1000 contiguous channels for simultaneous
$\lambda$$\sim$3 and 2\,mm observations. At $\lambda\sim$1.3 and
0.85\,mm, all 2000 channels were used to cover a similar velocity
range. The channel separation of 43\,kHz corresponds to
0.04--0.15\,km\,s$^{-1}$ for the frequency interval 357--85\,GHz.

All measurements were made with a circular rotating disk
to enable the beam to be either reflected or to pass through 
to a mirror system behind it. The two light beams allowed us 
to observe in a dual beam switching mode with a switching frequency 
of 6\,Hz (controlled by the rotational speed of the disk) 
and a beam throw of 11$'$40$''$ in azimuth. Since rapid beam 
switching was used in conjunction with reference positions on 
both sides of the source, baselines are of good quality. Calibration 
was obtained with the chopper wheel method. Main beam efficiencies 
of 0.746, 0.683, 0.457 and 0.30 at 94, 115, 230, and 345\,GHz, 
respectively, were derived from measurements of Jupiter (L. 
Knee, priv. comm.; see also Table~\ref{tel.param}). These 
values were interpolated and, if neccessary, also extrapolated
to convert antenna ($T_{\rm A}^{*}$) to main beam brightness
($T_{\rm mb}$) temperature. The pointing accuracy, obtained from
measurements of the nearby SiO maser source R Dor, was mostly better
than 10$''$ (see also Sect.\,3).

\subsubsection{APEX 12-m observations}

In October 2007, observations of the CO $J$=3--2 line were carried out 
with the double sideband APEX-2a facility receiver (Risacher et 
al.  2006). 145 positions were measured in a position switching mode 
with an on-source integration time of 40\,sec and a beamwidth of 
$\sim$20\arcsec\ (see Table~\ref{tel.param}). The beam efficiency 
was 0.73 and the forward hemisphere efficiency 0.97 (G{\"u}sten et 
al. 2006). Both units of a Fast Fourier Transform Spectrometer with 
1\,GHz bandwidth and 16384 channels each (Klein et al. 2006) were 
used to measure the CO transition. System temperatures on a 
$T_{\rm A}^*$ scale were 300--450\,K.

\section{Results}

\begin{figure}
\hspace{-28mm} \vspace{33mm}
\includegraphics[angle=-90,width=10cm]{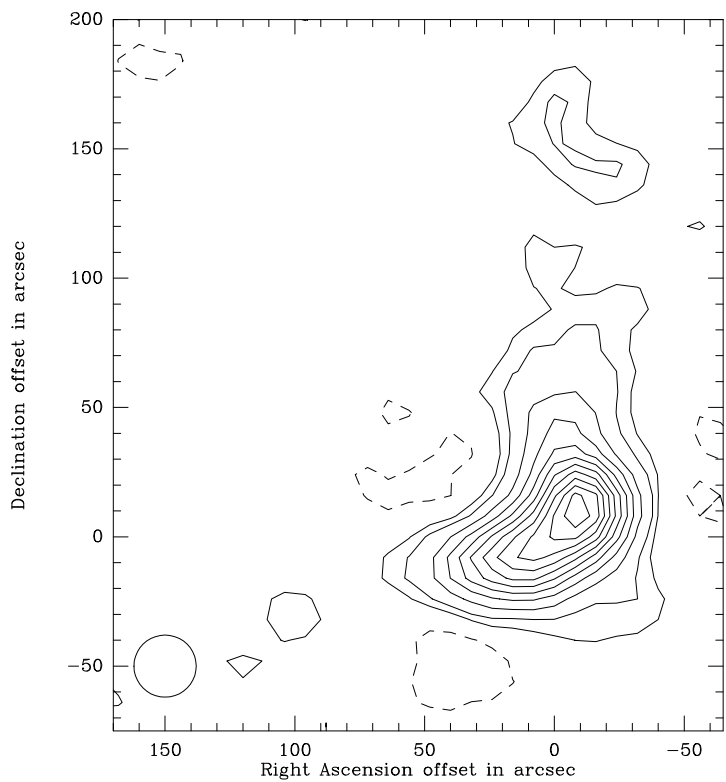}
\vspace{7mm}
\caption{A $\lambda$1.2\,mm continuum map of N113.
Contour levels are --1, 1, 2, 4, 6, 8, ... 20 times
19\,mJy\,beam$^{-1}$; the 1$\sigma$ noise level is
7\,mJy\,beam$^{-1}$. The peak is at 420\,mJy\,beam$^{-1}$. The
reference position is $\alpha_{\rm J2000}$ = 05$^{\rm h}$ 13$^{\rm
m}$ 18.$\!^{\rm s}$2, $\delta_{\rm J2000}$ = --69$^{\circ}$
22\arcmin\ 35\arcsec ($\alpha_{\rm B1950}$ = 05$^{\rm h}$ 13$^{\rm
m}$ 38.$\!^{\rm s}$7, $\delta_{\rm B1950}$ = --69$^{\circ}$
25\arcmin\ 57\arcsec). The calibration is estimated to be accurate
to $\pm$20\%. Pointing uncertainty and beam size (the circle in the
lower left corner) are $\pm$7\arcsec\ and 24\arcsec, respectively.
\label{n113.1}}
\end{figure}

\begin{figure}
\includegraphics[angle=-90,width=10cm]{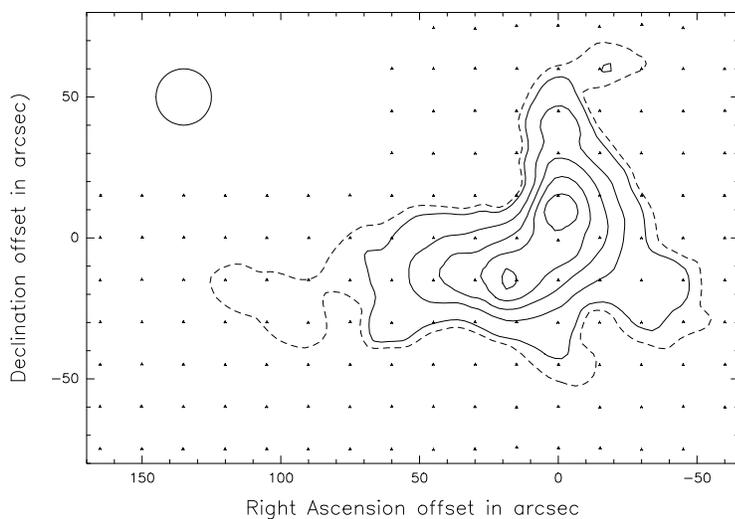}
\vspace{-6mm} \caption{An APEX CO $J$=3--2 map showing intensities
integrated over the Local Standard of Rest (LSR) velocity interval
225--245\,km\,s$^{-1}$. Contours are 7 (dashed), 10, 20, 30, 40, and
50\,K\,km\,s$^{-1}$ on a $T_{\rm A}^*$ scale. The reference position
is $\alpha_{\rm J2000}$ = 05$^{\rm h}$ 13$^{\rm m}$ 18.$\!^{\rm s}$2,
$\delta_{\rm J2000}$ = --69$^{\circ}$ 22\arcmin\ 35\arcsec. To
convert to scales of main beam brightness temperature, multiply by
1.35. The calibration is estimated to be accurate to $\pm$15\%.
Pointing uncertainty and beam size (the circle in the upper left
corner) are $\pm$3\arcsec\ and 20\arcsec, respectively.
\label{n113a.1}}
\end{figure}

Figure\,\ref{n113.1} shows a 1.2\,mm continuum map that is sensitive
to the column density of the interstellar dust. Outside the core of
N\,113, which shows an elongation along an axis extending from the
north-west to the south-east, we find protrusions toward the north
and east, the latter only slighty surpassing the noise level.
Figure\,\ref{n113a.1} shows the corresponding distribution of
integrated CO $J$=3--2 emisison. While the morphology appears to be
similar, the eastern tongue is detected with higher significance.
The dashed outer contour still represents a 6$\sigma$ level.

\begin{figure}
\includegraphics[angle=-90,width=17cm]{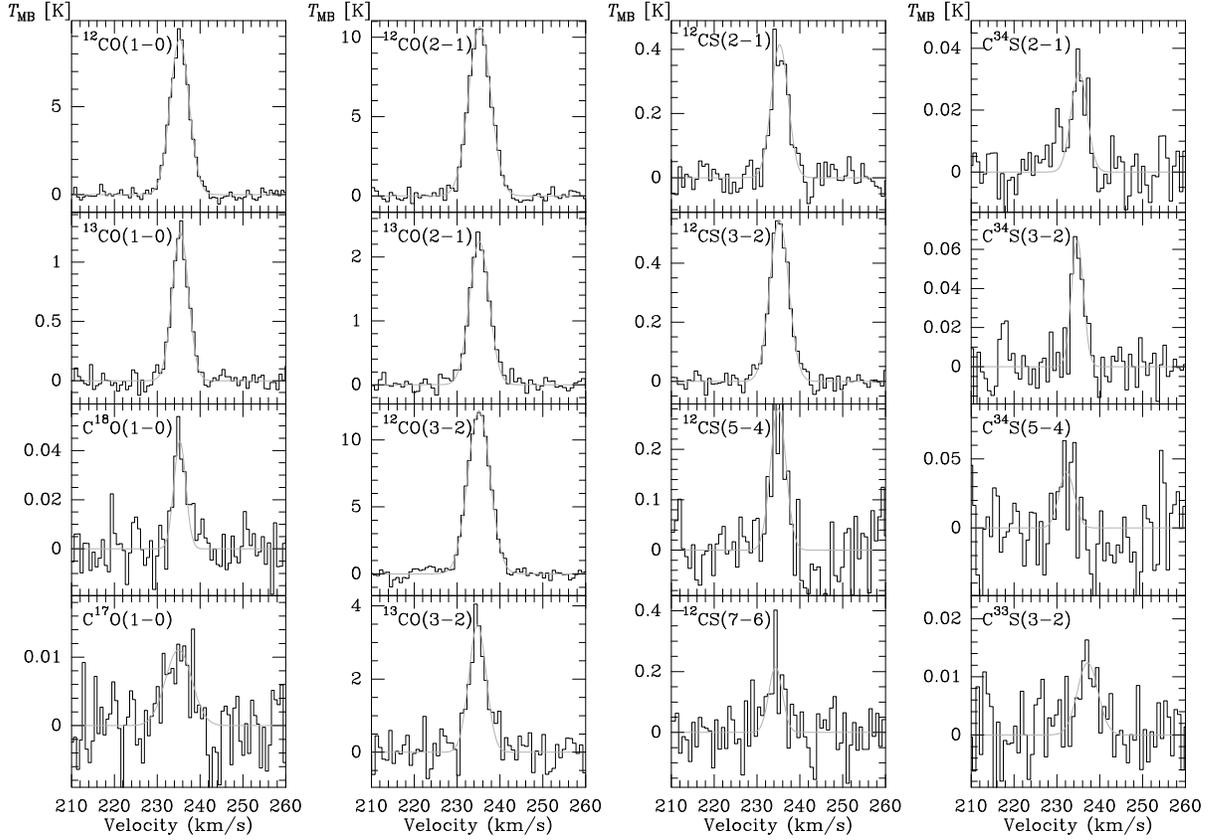}
\caption{CO and CS spectra measured toward N113 ($\alpha =05^{\rm
h}13^{\rm m}18.2^{\rm s}$, $\delta = -69^{\rm o}22' 35''$ (J2000)).
The channel spacing for each spectrum is $\sim\,$0.9\,km\,s$^{-1}$.
Gaussian fits were performed for detected and tentatively detected
lines, and resulting parameters are listed in Table~\ref{lin.para}
(same in Figs.\,\ref{n113.3}--\ref{n113.5}). \label{n113.2}}
\end{figure}

\begin{figure}
\includegraphics[angle=-90,width=17cm]{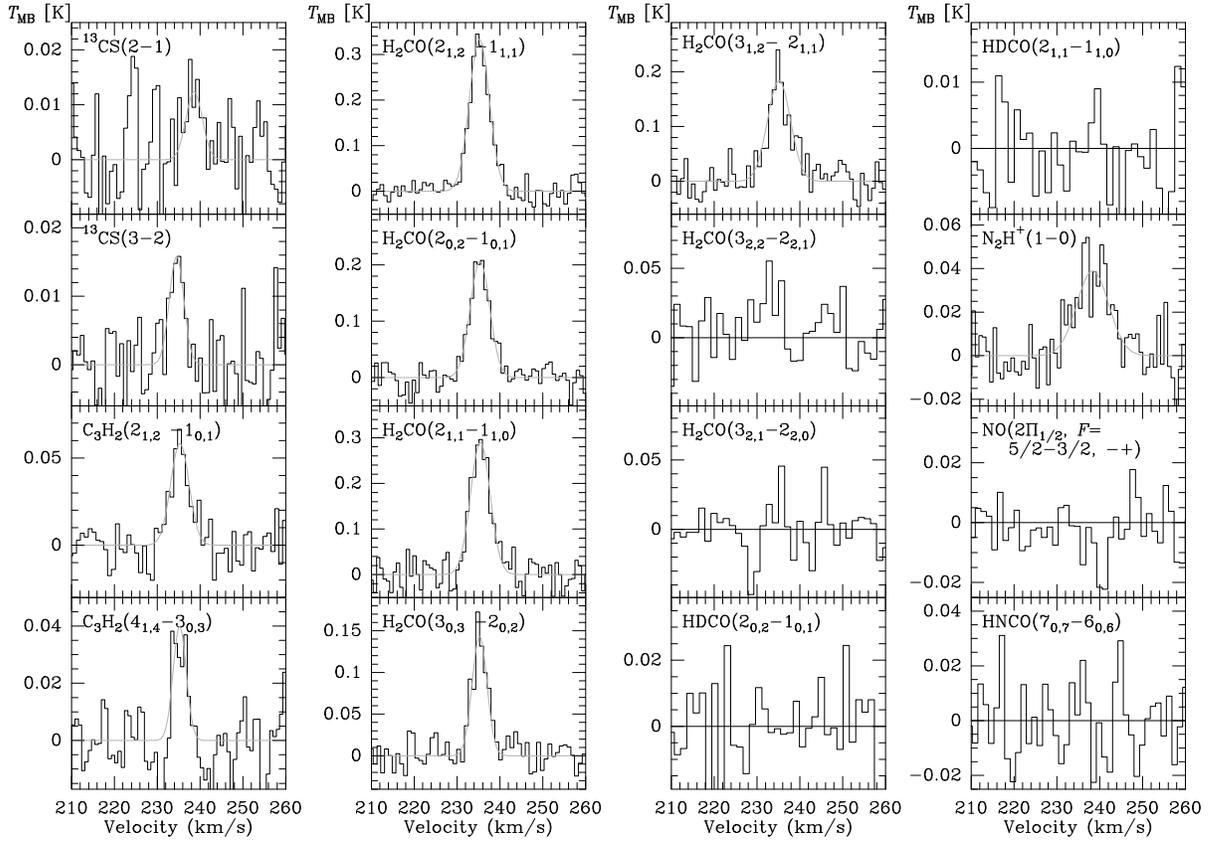}
\caption{CS, C$_3$H$_2$, H$_2$CO, HDCO, N$_2$H$^+$, NO and HNCO
spectra measured toward N113. In the case of N$_2$H$^+$, the 
frequency of the main hyperfine component (93.713809\,GHz) is
taken as reference. In Table~\ref{lin.para}, however, the expected 
average frequency for optically thin emission and Local Thermodynamical 
Equilibrium is displayed. For further details, see Fig.\,\ref{n113.2}.
\label{n113.3}}
\end{figure}

\begin{figure}
\includegraphics[angle=-90,width=17cm]{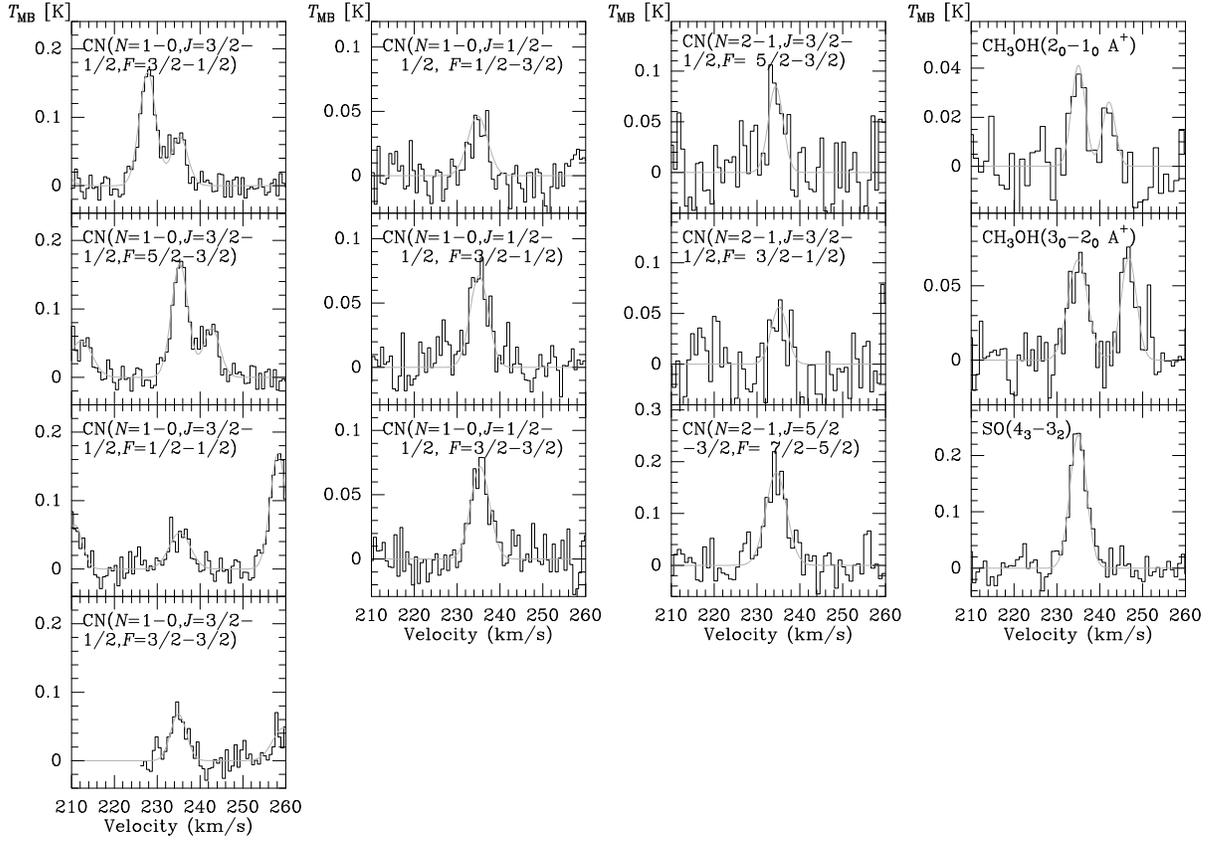}
\caption{CN, CH$_3$OH, and SO spectra measured toward N113. The
CH$_3$OH figures also show the 2$_{-1}$--1$_{-1}$ E and
3$_{-1}$--2$_{-1}$ E lines at slightly higher velocities than the
2$_0$--1$_0$ A$^+$ and 3$_0$--2$_0$ A$^+$ profiles, respectively.
For further details, see Fig.\,\ref{n113.2}. \label{n113.5}}
\end{figure}

Figures\,\ref{n113.2}--\ref{n113.5} display the spectra measured
toward the peak of N\,113. Agreement between the APEX and SEST CO
$J$=3--2 spectra at the central position is reasonably good (peak
line temperatures are 10.5 and 12\,K on a $T_{\rm mb}$ scale,
respectively). SEST line parameters are given in
Table~\ref{lin.para} and include 50 detected, 7 tentatively
detected, and 6 undetected transitions. The $^{12}$C$^{34}$S
$J$=5--4, $^{12}$C$^{33}$S 3--2, $^{13}$C$^{32}$S 2--1, and
HC$_3$N 10--9 spectral features show deviations from a radial
velocity of $\sim$235\,km\,s$^{-1}$. These are, however, likely 
caused by noise, as the emission in the above transitions is 
significantly weaker than in the other listed lines of CS. The 
relatively high velocity of the N$_2$H$^+$ 1--0 transition 
(Table~\ref{lin.para}) may be a consequence of the 
presence of seven hyperfine components. Deviations from the relative 
intensities expected in the case of Local Thermodynamic Equilibrium 
(LTE) could shift the line by $\sim$1--2\,km\,s$^{-1}$. Alternatively,
an exceptional drift in the velocity scale of the temperature 
sensitive backend cannot be excluded. Among the 16 molecules 
observed (this includes 28 ``isotopologues'', i.e. species 
containing different isotopic substitutions), only two remain 
undetected, NO and HCNO. Both contain nitrogen.

\begin{figure}
%\begin{minipage}[t]{6.5cm}
\centering
\includegraphics[width=.8\textwidth,angle=0]{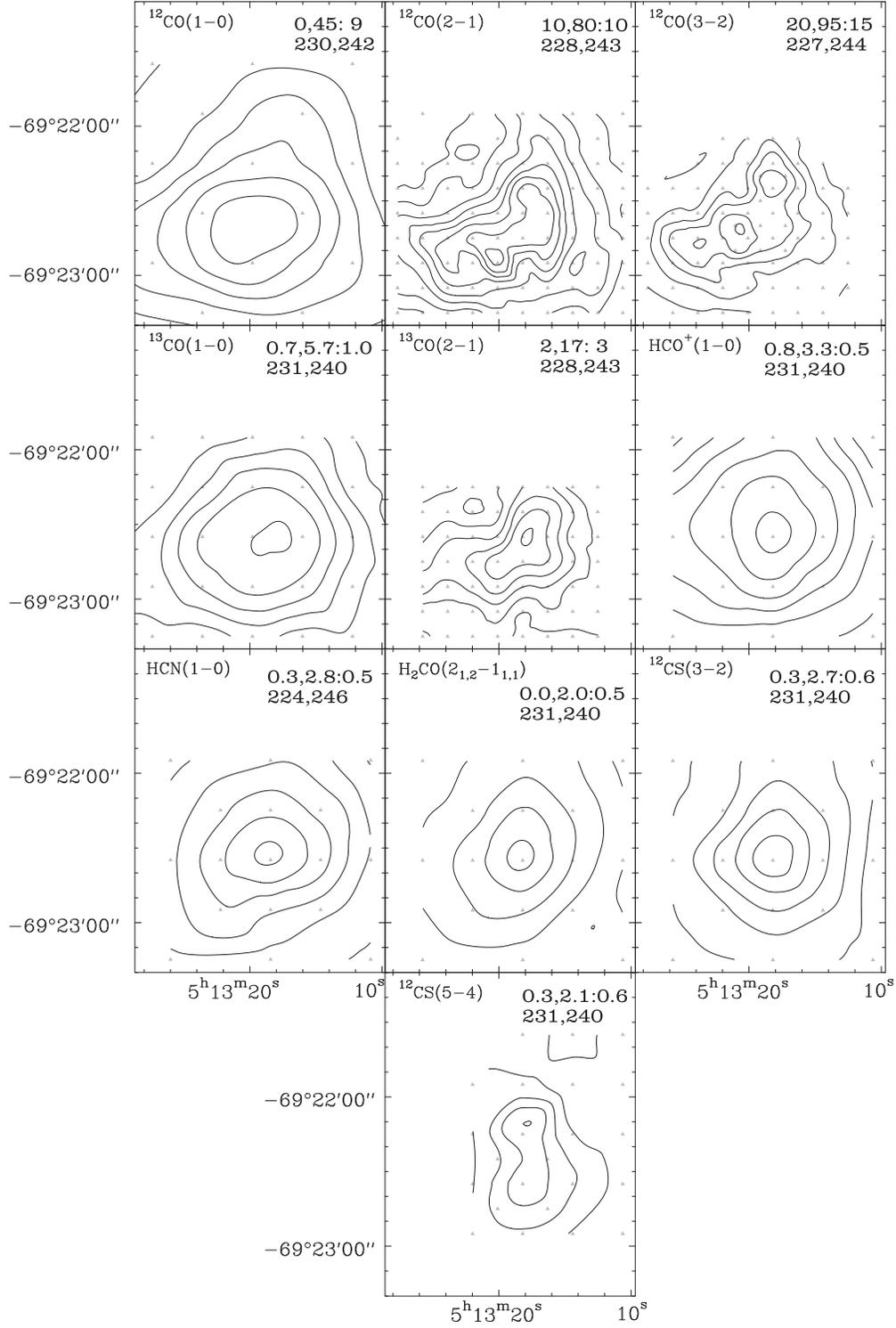}
\caption{SEST CO, HCO$^+$, HCN, H$_2$CO, and CS maps toward N113.
The upper left corner of each panel indicates the molecular
transition; the upper right corner gives the lowest contour,
the highest contour, and the increment in K\,km\,s$^{-1}$ (upper
line) and the velocity range in km\,s$^{-1}$ (lower line).}
%\end{minipage}
\label{n113.b1}
\end{figure}

Also obtained were small maps in 10 molecular lines, including CO,
$^{13}$CO, and the higher density tracers CS, HCN, HCO$^+$, and
H$_2$CO. Observed positions and contour plots are displayed in
Fig.\,\ref{n113.b1}. The maps outline to a certain degree the extent
of the molecular cloud. Since its size is, however, often comparable
to the size of the beam of the telescope, a deconvolution of the
beam was necessary. The results of this deconvolution are given in
Table~\ref{cloudsize}. Shown are the transition (Cols.\,1 and 2),
the beam width (Col.\,3), the observed Full Width to Half Power
(FWHP) source size in right ascension and declination (Cols.\,4 and
5), the deconvolved extent in right ascension and declination
(Cols.\,6 and 7) and its geometric mean (Col.\,8). In view of map
spacing and possible pointing errors, resulting intrinsic source
sizes are not acccurate. Thus for CO $J$=3--2 SEST
(Fig.\,\ref{n113.b1}) and APEX (Fig.\,\ref{n113a.1}) data
deconvolved source sizes are 40 and 60\arcsec, respectively. For
HCN, the intrinsic size of the emitting region remains undetermined,
while the half power extent of the $\lambda$1.2\,mm continuum
emission appears to be intermediate between those of CO $J$=1--0 and
2--1 and the average value derived from the molecular high density
tracers (see Sect.\,4.2 for adopted source sizes).

\section{Discussion}

N\,113 is with N\,159 one of the two strongest molecular line 
emitters of the Magellanic Clouds. Prior to a detailed analysis of
our spectra, some general source properties should be mentioned. 
From the IRAS flux densities (5.7, 40.5, 268, and 415\,Jy at 12.5, 
25, 60, and 100$\mu$m, respectively) we obtain a total luminosity of 
$\sim$2$\times$10$^6$\,L$_{\odot}$ for N\,113, extrapolating from 6
to 400$\mu$m and assuming a grain emissivity proportional to $\nu$
(see Wouterloot \& Walmsley 1986). Wong et al. (2006) collected
$\lambda$3\,mm SEST, Mopra, and ATCA (Australia Telescope Compact
Array) data of the radio continuum and prominent molecular species
toward N\,113. While the 3\,mm continuum only reveals a flux density
of $\sim$40\,mJy from a 5--10\arcsec\ sized source (presumably,
there is missing flux), their SEST HCN and HCO$^+$ spectra toward
the cloud core are consistent with our data. In agreement with our
measurements (Fig.\,\ref{n113.b1}, Table~\ref{cloudsize}), their
SEST and Mopra maps also show a source that is only slightly more
extended than the telescope beam. The interferometric high
resolution ATCA maps did not collect the entire flux but reveal a
compact cloud core of size 8\arcsec$\times$5\arcsec, with HCO$^+$
$J$=1--0 being presumably more extended than HCN $J$=1--0.

In the following we first discuss the $\lambda$1.2\,mm continuum
(Sect.\,4.1) and then proceed to the analysis of spectral lines from
individual molecular species (Sects.\,4.2 and 4.3). Summaries of the
observed properties with respect to H$_2$ densities (Sect.\,4.4),
molecular column densities (Sect.\,4.5), and stellar 
nucleosynthesis (Sect.\,4.6) follow.

\subsection{The dust continuum}

The 60 and 100$\mu$m IRAS fluxes and an emissivity proportional to
$\nu$ yield a dust color temperature of $T_{\rm d(60/100)}$ =
42.5\,K. With an integrated $\lambda$1.2\,mm flux density of 1.6\,Jy
(see Fig.\,\ref{n113.1}) and an emissivity proportional to $\nu^2$
we obtain a 100$\mu$m to 1.2\,mm dust color temperature $T_{\rm
d(100/1200)}$ = 21\,K. The uncertainty in the 1.2\,mm flux density
of $\pm$20\% causes an error in $T_{\rm dust}$ that does not surpass
$\pm$2\,K. Applying the equation given in Table~\ref{cloudsize} of
Mauersberger et al. (1996a), we then obtain for dust temperatures
between 20 to 45\,K a total gas mass of $M_{\rm N113}$ = (6.7 --
3.0)$\times$10$^5$\,M$_{\odot}$. This assumes a dust to gas mass
ratio of 500, which is a factor of four higher than in the solar 
neighborhood (Bolatto et al. 2000). For the 24$''$ sized central 
area (420\,mJy), the column density becomes 
(1.0 -- 0.5)$\times$10$^{23}$\,cm$^{-2}$ for the range of plausible 
dust temperatures (see Sect.\,4.2.1 for the corresponding estimate 
from CO). The total mass estimated by us is a little larger than 
that given by Wong et al. (2006) who favored 
$M_{\rm N113}$$\sim$10$^5$\,M$_{\odot}$.

\subsection{Molecules with LVG modelling}

With a Large Velocity Gradient (LVG) model (e.g., Sobolev 1960;
Castor 1970; Scoville \& Solomon 1974) and choosing a spherically
symmetric cloud geometry, the H$_2$ density and the column density
of a given species can be estimated. As input this requires some
knowledge of the kinetic temperature as well as line intensities
from a sufficient number of transitions of a given molecule. The
choice of a particular cloud geometry can affect resulting densities
by up to half an order of magnitude, but only if the lines are
optically thick. Applying a plane-parallel instead of a spherical
cloud geometry can result in particle densities which are lower by
up to this amount.

We correct for beam dilution by calculating $T^{\prime}_{\rm mb}$ =
$T_{\rm mb}$/$\eta_{\rm bf}$ with $\eta_{\rm bf}$ = $\theta^2_{\rm
s}$/($\theta^2_{\rm s}$+$\theta^2_{\rm b}$). $\theta_{\rm b}$ and
$\theta_{\rm s}$ denote beam and source size, respectively (see also
Wang et al. 2004). With the exception of a few CO lines (see
Sect.\,3) a source size of 40$''$ was assumed, which is consistent
with the extent of the 1.2\,mm continuum emission (Sect.\,3,
Fig.\,\ref{n113.1}, and Table~\ref{cloudsize}) and with the average
of the (individually uncertain) molecular cloud size estimates of
Table~\ref{cloudsize}. Table~\ref{lin.int} shows line temperatures 
prior and after correction for beam dilution.

\begin{figure}
%\begin{minipage}[t] {18.5cm}
\includegraphics[angle=-90,width=15cm]{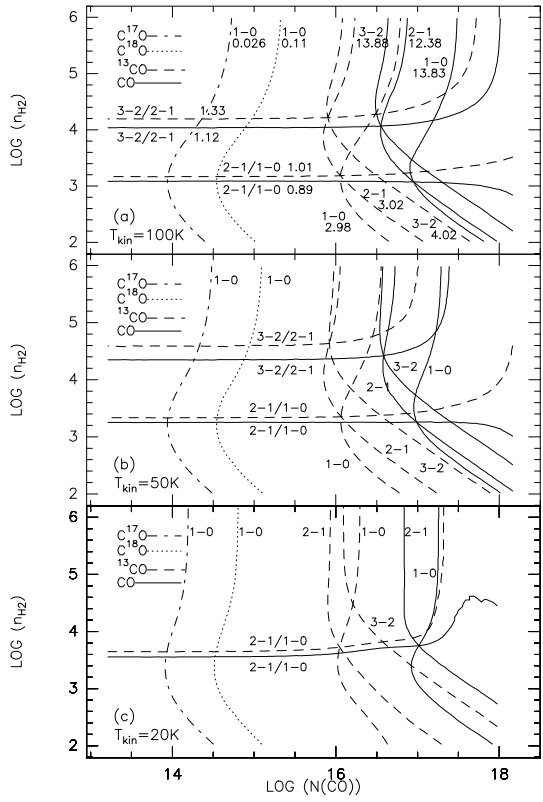}
%\end{minipage}
\caption{LVG calculations for C$^{17}$O (dash-dotted lines),
C$^{18}$O (dotted lines), $^{13}$CO (dashed lines) and CO (solid
lines) at $T_{\rm kin}$\,$\sim$\,100\,K (a), $T_{\rm
kin}$\,$\sim$\,50\,K (b) and $T_{\rm kin}$\,$\sim$\,20\,K (c).
\label{n113.c1}}
\end{figure}

\begin{figure}
%\begin{minipage}[t] {18.5cm}
\includegraphics[angle=-90,width=15cm]{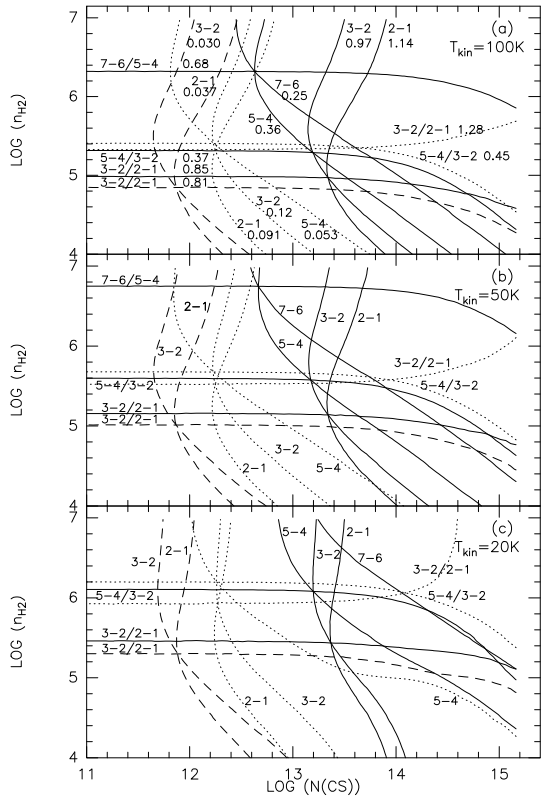}
%\end{minipage}
\caption{LVG calculations for CS (solid lines) and its rare
isotopologues C$^{34}$S (dotted lines) and $^{13}$CS (dashed lines)
at $T_{\rm kin}$\,$\sim$\,100\,K (a), $T_{\rm kin}$\,$\sim$\,50\,K
(b) and $T_{\rm kin}$\,$\sim$\,20\,K (c). \label{n113.c2}}
\end{figure}

\begin{figure}
%\begin{minipage}[t] {18.5cm}
\includegraphics[angle=-90,width=15cm]{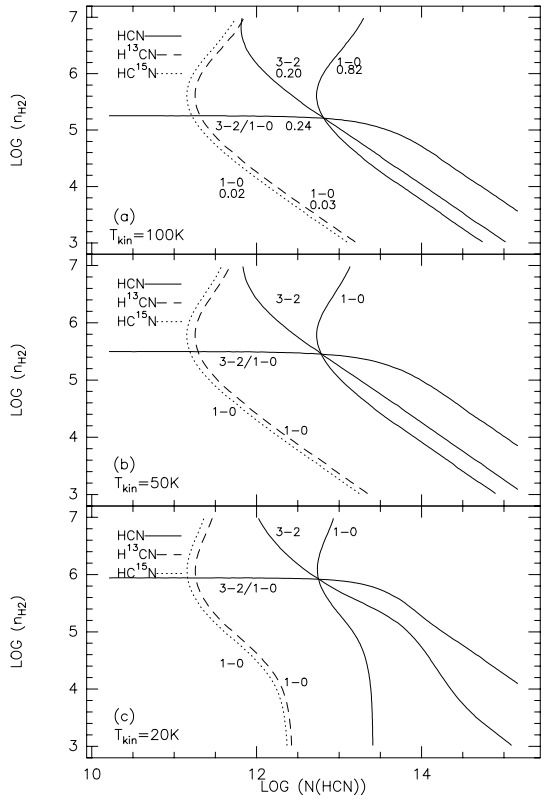}
%\end{minipage}
\caption{LVG calculations for HCN and its rare isotopologues
H$^{13}$CN and HC$^{15}$N at $T_{\rm kin}$\,$\sim$\,100\,K (a),
$T_{\rm kin}$\,$\sim$\,50\,K (b) and $T_{\rm kin}$\,$\sim$\,20\,K
(c). \label{n113.c3}}
\end{figure}

\begin{figure}
%\begin{minipage}[t] {18.5cm}
\includegraphics[angle=-90,width=15cm]{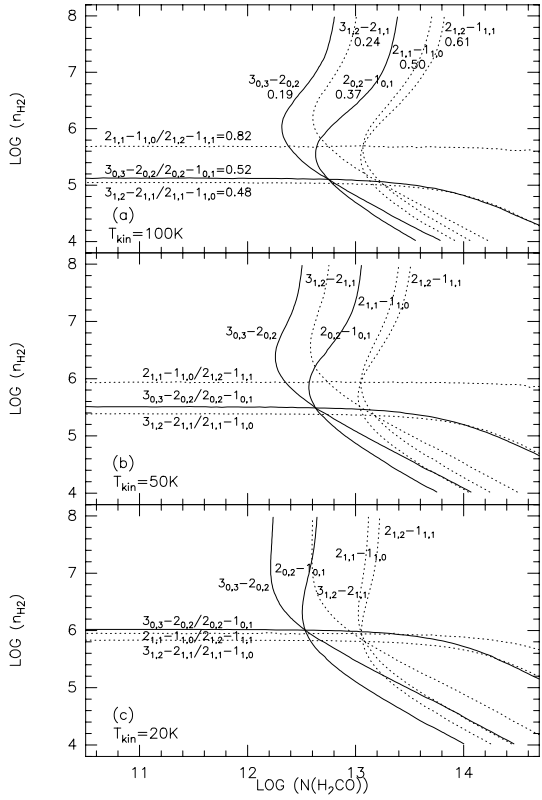}
%\end{minipage}
\caption{LVG calculations for para-H$_2$CO (indicated by solid
lines) and ortho-H$_2$CO (dotted lines) at $T_{\rm kin}$ = 100\,K
(a), $T_{\rm kin}$ = 50\,K (b) and $T_{\rm kin}$ = 20\,K (c).
\label{n113.c5}}
\end{figure}

\begin{figure}
%\begin{minipage}[t] {18.5cm}
\includegraphics[angle=-90,width=15cm]{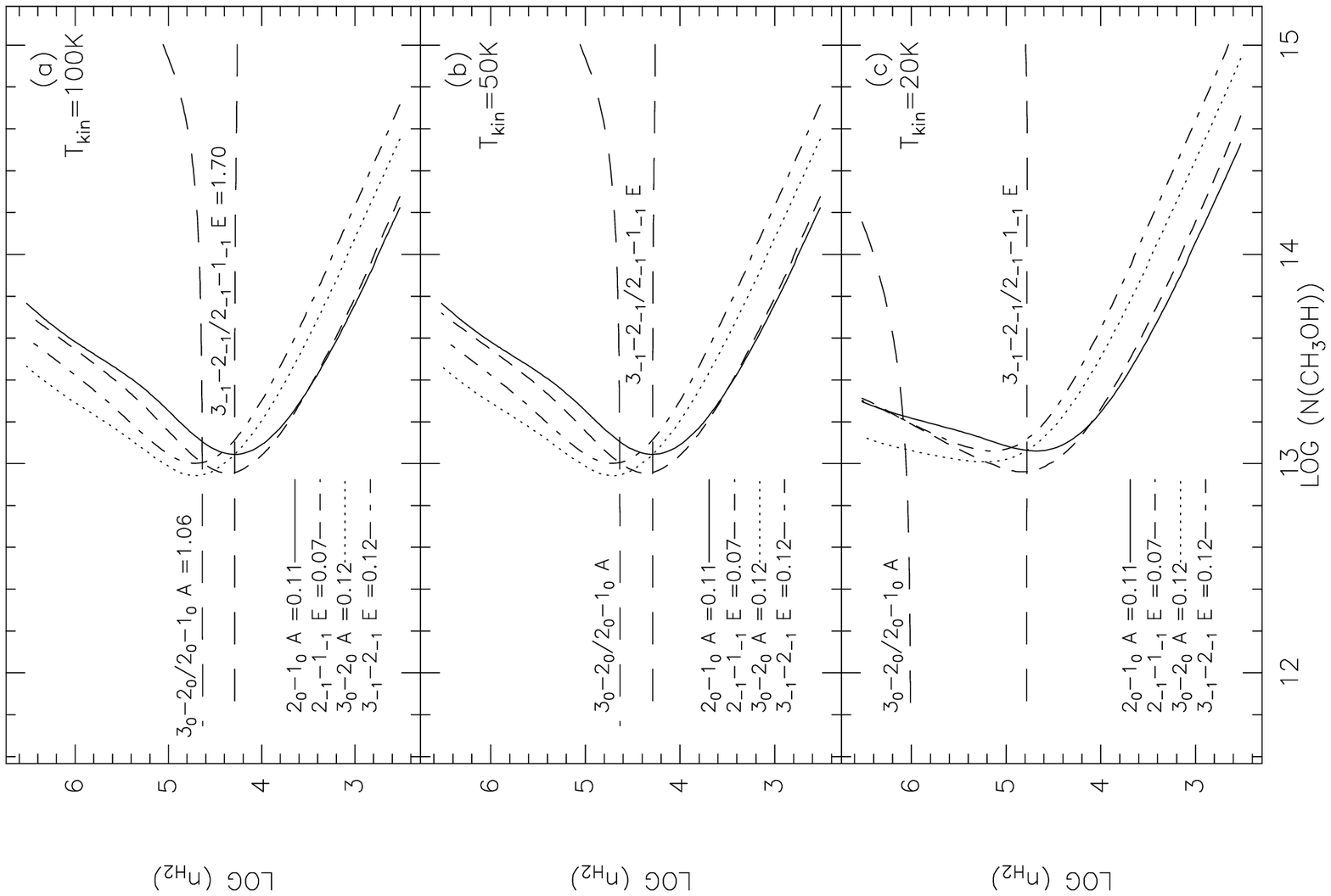}
%\end{minipage}
\caption{LVG calculations for CH$_3$OH at $T_{\rm
kin}$\,$\sim$\,100\,K (a), $T_{\rm kin}$\,$\sim$\,50\,K (b) and
$T_{\rm kin}$\,$\sim$\,20\,K (c). \label{n113.c6}}
\end{figure}

The results of the model calculations that are based on the
corrected line temperatures are shown in
Figs.\,\ref{n113.c1}--\ref{n113.c6}. Displayed are line intensities
and peak line intensity ratios as a function of H$_2$ density and
molecular column density.

In the following, LVG simulations are discussed for the most 
important molecular species. Calculations were made for
kinetic temperatures of 20, 50, and 100\,K (see e.g., Mangum \&
Wootten 1993, Chin et al. 1996b, and Heikkil{\"a}, et al. 1999 for
a justification of the chosen temperature range).

\subsubsection{CO}

All of the eight lines of carbon monoxide (CO) shown in the left
panels of Fig.\,\ref{n113.1} have been clearly detected. The
lineshapes are compatible. While the $^{12}$C$^{17}$O (hereafter
C$^{17}$O) $J$=1--0 line is broader than all other measured CO
transitions, this is not an effect caused by a relatively low
signal-to-noise level. Hyperfine splitting, leading to two blended
main spectral features separated by a few km\,s$^{-1}$ (see Lovas
\& Krupenie 1974; Wouterloot et al. 2005), is broadening the line.
A comparison of the $^{12}$C$^{18}$O (hereafter C$^{18}$O) and
C$^{17}$O $J$=1--0 lines results in an intensity ratio of
1.7$\pm$0.2 which is extremely low compared with galactic values
measured by Penzias (1981) and Wouterloot et al. (2005). The line
intensity ratio agrees, however, with the average ratio of
1.6$\pm$0.3 determined by Heikkil{\"a} et al. (1999) from the
$J$=2--1 lines for a number of prominent H{\sc ii} regions of the
LMC. As a consequence it is reasonable to assume that C$^{18}$O
and C$^{17}$O are optically thin and that potential differences in
shielding (Heikkil{\"a} et al. 1999) are minimal (see Sect.\,4.6).

Because of optical depth effects, an interpretation of the
$^{12}$C$^{16}$O (hereafter CO) and $^{13}$C$^{16}$O (hereafter
$^{13}$CO) spectra is less straightforward. With a
$^{12}$C/$^{13}$C isotope ratio of 49$\pm$5 (Sect.\,4.2.3) and
$^{12}$CO/$^{13}$CO line intensity ratios of 7.1, 4.8 and 3.5 for
the $J$=1--0, 2--1 and 3--2 transitions, respectively (see
Tables~\ref{lin.para} and \ref{tab.den}), CO opacities must be of
order $\tau$ $\sim$ 7, 10, and 14. These are consistent with
opacities of typical galactic clouds (e.g. Larson 1981), but are
much larger than $\tau$$\sim$1, which was proposed by Heikkil{\"a}
et al. (1999) for the CO $J$=1--0 transition in other star forming
regions of the LMC. For our estimate we have assumed that 
CO and $^{13}$CO excitation temperatures are similar and that
fractionation (Watson et al. 1976) and isotope selective
photodissociation (Bally \& Langer 1982) are either not important
or are balancing each other. For the Galaxy, prominent molecular
clouds do not show strong differences in the relative abundances
of $^{12}$C and $^{13}$C bearing isotopologues (for CO, CN, and
H$_2$CO, see Milam et al. 2005). For the LMC this is less clear
(e.g., Heikkil{\"a} et al. 1999). In any case, the measured
CO/$^{13}$CO line intensity ratios are large enough to ensure that
$^{13}$CO is optically thin. Larger opacities resulting in the
observed smaller $^{12}$C/$^{13}$C line intensity ratios in the
higher--$J$ rotational CO transitions are expected because of the
higher statistical weights of the molecular states involved.

While $^{18}$O/$^{17}$O and $^{12}$C/$^{13}$C isotope ratios have
previously been determined in several star forming regions of the
LMC and while the gas appears to be well mixed, showing similar
ratios in different sources (see Heikkil{\"a} et al. 1998 for the
oxygen ratio), the higher and thus more elusive $^{16}$O/$^{18}$O
isotope ratio was so far only determined in N\,159W (Heikkil{\"a}
et al. 1999, their Table~16). With the $^{12}$C/$^{13}$C isotope
ratio directly obtained from the relative strengths of the HCN
$J$=1--0 hyperfine components and the line intensity ratio of its
$^{12}$C and $^{13}$C bearing species (Sect.\,4.2.3 and Chin et
al. 1999), a more accurate estimate is possible for N\,113. If
$^{13}$CO $J$=1--0 is optically thin as indicated above, the
$^{16}$O/$^{18}$O ratio is determined by multiplying the
$^{13}$CO/C$^{18}$O $J$=1--0 line intensity ratio by 49$\pm$5,
which is the $^{12}$C/$^{13}$C isotope ratio deduced from HCN
(Sect.\,4.2.3). We obtain $^{16}$O/$^{18}$O = 2000$\pm$250. This
result is in good agreement with the value given in Table~16 of
Heikkil{\"a} et al. (1999) for N\,159W, suggesting that the ratio,
like that of $^{18}$O/$^{17}$O, may not vary strongly from source
to source. The $^{16}$O/$^{18}$O value is four to ten times larger
than corresponding ratios determined in the interstellar medium of
spiral galaxies (e.g., Henkel \& Mauersberger 1993). It yields a
double isotope abundance ratio of $^{13}$CO/C$^{18}$O $\sim$ 40,
also consistent with the value proposed by Heikkil{\"a} et al.
(1998) for N\,159W. For $^{16}$O/$^{17}$O, we obtain 3400$\pm$600.

With $^{12}$CO opacities of order 10, a carbon isotope ratio of
$\sim$50, and the $^{16}$O/$^{18}$O ratio determined above,
$^{13}$CO should be optically thin not only in the $J$=1--0, but
also in the 2--1 and 3--2 transitions. Multiplying the $^{13}$CO
column density derived from LVG modeling by 50, taking collision
rates with H$_2$ from Flower (2001), and assuming an 
ortho-to-para H$_2$ abundance ratio of 3:1 allows us to determine 
the total CO column density. With the LVG code we obtain
$N$(CO)$\sim$6.5$\times$10$^{17}$\,cm$^{-2}$ and $n$(H$_2)$ $\sim$
5$\times$10$^{3}$\,cm$^{-3}$. For a graphic display of the results 
that neither strongly depend on the choice of the collision rates 
nor on the ortho-to-para H$_2$ abundance ratio, see Fig.\,\ref{n113.b1}. 
The resulting values hold approximately for kinetic temperatures 
between 20 and 100\,K.

Applying the mass estimate given by MacLaren et al. (1988) for a
virialized cloud with an 1/$r$ density gradient, we find for a
cloud size of $R$=5.5\,pc (23\arcsec, Table~\ref{tel.param}; this
is within the range of beam radii listed in Table~\ref{cloudsize})
and a linewidth of 5\,km\,s$^{-1}$, $N$(H$_2$) $\sim$
1.7$\times$10$^{22}$\,cm$^{-2}$, $N$(CO)/$N$(H$_2$) $\sim$
4$\times$10$^{-5}$, and $<$$n$(H$_2$)$>$$\sim$500\,cm$^{-3}$. The
CO/H$_2$ abundance ratio is about half that characterizing
galactic clouds (e.g., Frerking et al. 1982). The conversion
factor between H$_2$ column density and CO $J$=1--0 integrated
intensity becomes with $I_{\rm CO~1-0}$ $\sim$ 50\,K\,km\,s$^{-1}$
(Table~\ref{lin.para}) $X$ = $N$(H$_2$)/$I_{\rm CO~1-0}$ $\sim$
3.4$\times$10$^{20}$\,cm$^{-2}$\,(K\,km\,s$^{-1}$)$^{-1}$. 
This is almost twice the value for the galactic disk (e.g.,
Mauersberger et al. 1996a, their Appendix A.1) and the value
reported by Chin et al. (1997) on the basis of a small number of 
CO lines observed toward N\,113. Since uncertainties amount to 
at least a factor of two, our value is still consistent with 
approximately galactic values on small spatial scales, as already 
suggested by Rubio et al. (1993) and Chin et al. (1997) for the 
Magellanic Clouds. The column density derived for a virialized 
cloud is smaller than that obtained from the dust continuum in 
Sect.\,4.1. This may in part be caused by the different beams 
considered. If most of the dust emission would arise from the 
compact core seen by Wong et al. (2006), we had to multiply the 
$N$(H$_2$) = (5--10)$\times$10$^{22}$\,cm$^{-2}$ from the dust 
emission by (24\arcsec/45\arcsec)$^2$, yielding column densities 
of (1.4--2.8)$\times$10$^{22}$\,cm$^{-2}$. This agrees well with 
the virial estimate from the CO data.

\subsubsection{CS}

Ten lines of carbon monosulfide (CS) were observed. The results are
shown in the right two panels of Fig.\,\ref{n113.2} and in the left
panel of Fig.\,\ref{n113.3}. While the three lower--$J$ lines of the
main isotopic species, $^{12}$C$^{32}$S (hereafter CS), are clearly
seen, the $J$=7--6 line is only tentatively detected. The rare
isotopologue $^{12}$C$^{34}$S (hereafter C$^{34}$S) was detected in
two and tentatively also in a third transition, while weak features
are also seen at the frequencies of the $J$=2--1 and 3--2 lines of
$^{13}$C$^{32}$S (hereafter $^{13}$CS) and of the $J$=3--2 line of
$^{12}$C$^{33}$S (hereafter C$^{33}$S).

Are the lines of the main CS species optically thick, like those
of CO? The $J$=3--2 transition, which is strongest, measured in
four isotopologues, provides useful hints. For the CS/$^{13}$CS
$J$=3--2 line intensity ratio we obtain 37$\pm$5, which is a
little less than the $^{12}$C/$^{13}$C isotope ratio of $\sim$50
(Sect.\,4.2.3). This implies that optical depths are not large,
minimizing effects of isotope selective photodissociation.
Deviations in the abundance ratio between the $^{12}$C and
$^{13}$C bearing species of CS and HCN should also be small (Langer 
et al. 1984). Therefore neglecting these two effects, we obtain 
an optical depth $\tau$ $\sim$ 0.65 for the $J$=3-2 line 
of the main species, if CS and $^{13}$CS excitation temperatures are
similar. For CS/C$^{34}$S $J$=3--2, the line intensity ratio is
11.3$\pm$1.1. This is about a factor of two lower than the
abundance ratio measured towards N\,159W (Heikkil{\"a} et al.
1999; their Table~16), again suggesting a moderate degree of
saturation in the CS $J$=3--2 line. With $\tau$(CS 3--2)$\sim$0.65
as derived above, the $^{32}$S/$^{34}$S sulfur isotope ratio
becomes $\sim$15, slightly lower than values in the Galaxy (Chin
et al. 1996a) and in N159W. If the tentatively detected C$^{33}$S
$J$=3--2 line is really almost as strong as the $^{13}$CS $J$=3--2
line, this would imply $^{32}$S/$^{33}$S $<$ 100. This value is
smaller than the ratio of 120--150 found in the solar system, the
local interstellar medium, and the late-type carbon star IRC+10216
(Mauersberger et al. 2004).

Because C$^{34}$S is certainly optically thin and exhibits stronger
lines than the other rare CS species, it is the isotopologue of
choice to simulate CS excitation and to determine CS column density
and H$_2$ density. LVG calculations (collision rates from Turner et
al. 1992) yield, multiplying the C$^{34}$S column by a factor of 15,
$N$(CS) $\sim$ 2.5$\times$10$^{13}$\,cm$^{-2}$ and $n$(H$_2$) $\sim$
10$^{6}$\,cm$^{-3}$ for kinetic temperatures between 20 and 100\,K.
With an H$_2$ density several orders of magnitude higher than that
determined for CO, CS must trace a different kind of molecular gas.

\subsubsection{HCN, HCO$^{+}$ and HNC}

\begin{figure}
\includegraphics[angle=-90,width=17cm]{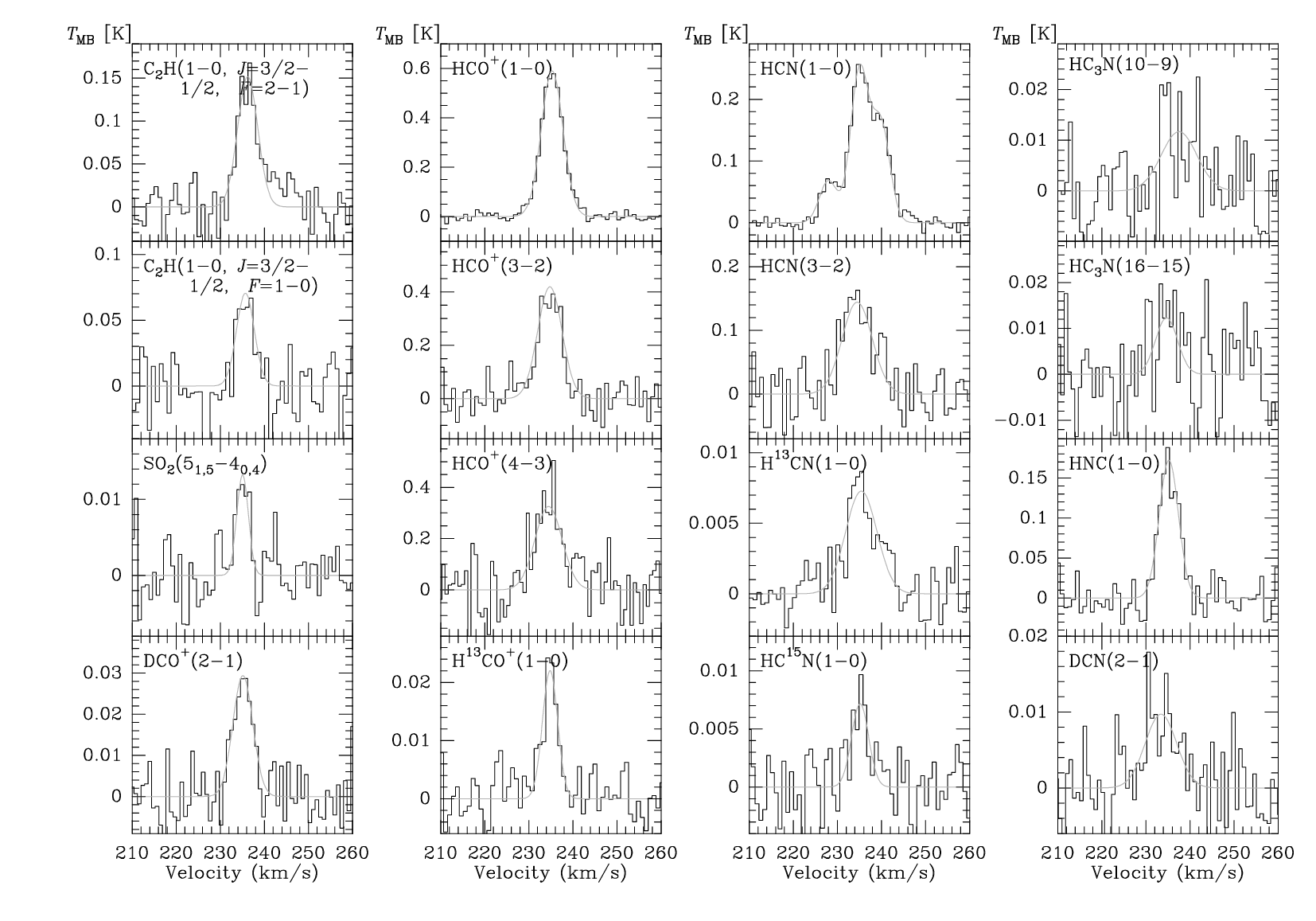}
\caption{C$_2$H, SO$_2$, DCO$^+$, HCO$^+$, HCN, HC$_3$N, HNC, and
DCN spectra measured toward N113. For details, see
Fig.\,\ref{n113.2}. \label{n113.4}}
\end{figure}

HCN is detected in the $J$=1--0 and 3--2 transitions
(Fig.\,\ref{n113.4}) together with the rare isotopologues
H$^{13}$CN and HC$^{15}$N, which were observed in the $J$=1--0
line, and DCN, which was measured in the $J$=2--1 line. For
earlier discussions of isotope ratios, see Chin et al. (1996b,
1999). Here we emphasize that the HCN $J$=1--0 line is
sufficiently narrow to show its hyperfine (HF) structure, i.e.
three components with relative intensities of 5:3:1 in the
optically thin limit under conditions of Local Thermodynamical
Equilibrium (LTE). The ratios actually observed are 3.64:2.34:1
(errors are of order 10\%), indicating a moderate degree of
saturation in its stronger HF components. Comparing the intensity
ratio between the weakest HF component ($\tau$$\sim$0.12, see
Table~2 of Chin et al. 1999) and the H$^{13}$CN line and
multiplying this by the factor (5+3+1)=9 yields a
$^{12}$C/$^{13}$C isotope ratio of 49$\pm$5. This is presumably
the most accurate carbon isotope ratio determined for the LMC
because we were able to derive the ratio from optically thin lines
(see, e.g., Johansson et al. 1994; Chin et al. 1996b;
Heikkil{\"a} et al. 1999 for consistent but less accurate 
determinations in other star forming regions of the LMC). 
According to Langer et al. (1984), the ratio from HCN may be 
somewhat higher than the overall carbon isotope ratio, but 
in the Galaxy, such differences are found to be negligible
(Milam et al. 2005). The ratio is thus used throughout the 
article.  For details on $^{14}$N/$^{15}$N and D/H, see Chin 
et al. (1996b, 1999).

Application of an LVG code (Fig.\,\ref{n113.c3}; collision rates
from Sch{\"o}ier et al. 2005) yields H$_2$ densities of
8$\times$10$^5$, 3$\times$10$^{5}$, and 1.6$\times$10$^5$\,cm$^{-3}$
for kinetic temperatures of 20, 50, and 100\,K. The column density
is $N$(HCN)\,$\sim$\,6.3$\times$10$^{12}$\,cm$^{-2}$.

HCO$^+$ was detected in a total of five lines, in the $J$=1--0, 3--2
and 4--3 transitions of the main species, the $J$=1--0 transition of
H$^{13}$CO$^+$ and the $J$=2--1 transition of DCO$^+$ (first and
second panel of Fig.\,\ref{n113.4}). For the $J$=1--0
HCO$^+$/H$^{13}$CO$^+$ line intensity ratio, we find 39$\pm$5, again
suggesting a moderate optical depth of order 0.5 as in the case of
the CS $J$=3--2 and HCN $J$=1--0 lines. LVG calculations
(Fig.\,\ref{n113.c4}) indicate H$_2$ densities of 6$\times$10$^{5}$,
2.5$\times$10$^{5}$, and 1.6$\times$10$^{5}$\,cm$^{-3}$ and a column
density of order 4$\times$10$^{12}$\,cm$^{-2}$. Note that for
$T_{\rm kin}$ = 20\,K, the $J$=4--3/$J$=3--2 line intensity ratio
cannot be reproduced, possibly suggesting that the kinetic
temperature of the gas is higher in this star forming region. This
induces a high uncertainty in the density estimate for this low
kinetic temperature.

Because a rare HNC isotopologue has not been observed, the optical
depth of its $J$=1--0 lines remains undetermined. Nevertheless,
because the line is weaker than those of HCN and HCO$^+$, the
assumption of optically thin emission appears to be reasonable.
Physical parameters of HCN and HNC are quite similar but chemical
properties differ strongly (e.g., Schilke et al. 1992; Aalto et al.
2002). In spite of this assuming similar spatial distributions and
excitation conditions, we tentatively obtain a total molecular
column density of $N$(HNC) $\sim$ 2.5$\times$10$^{12}$\,cm$^{-2}$.

\subsubsection{H$_2$CO}

\begin{figure}
%\begin{minipage}[t] {18.5cm}
\includegraphics[angle=-90,width=15cm]{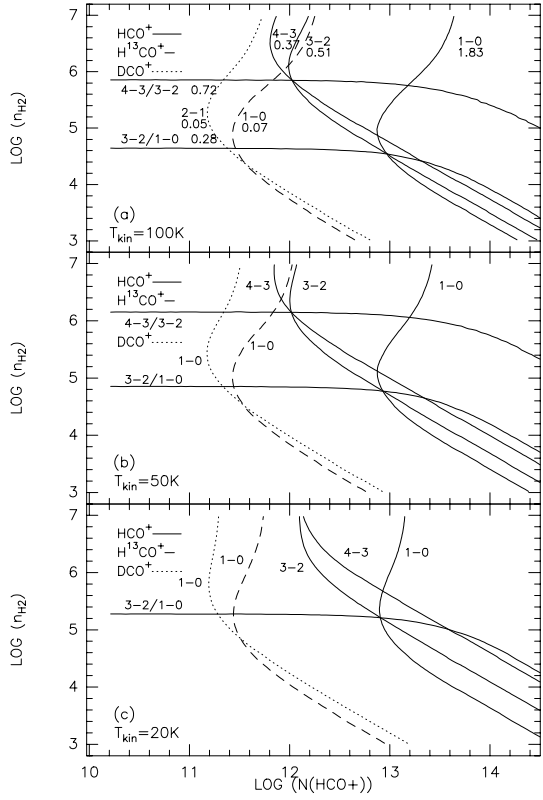}
%\end{minipage}
\caption[]{LVG calculations for HCO$^+$ and its rare isotopologues
H$^{13}$CN and DCO$^+$ at $T_{\rm kin}$\,$\sim$\,100\,K (a), $T_{\rm
kin}$\,$\sim$\,50\,K (b) and $T_{\rm kin}$\,$\sim$\,20\,K (c). Note
that for $T_{\rm kin}$ = 20\,K, no solution is found for the
$J$=4--3/$J$=3--2 line intensity ratio. \label{n113.c4}}
\end{figure}

At least five lines of formaldehyde (H$_2$CO) are seen (second and
third panels of Fig.\,\ref{n113.3}). Three belong with $K_{\rm
a}$=1 to the ortho-species (the hydrogen atoms have parallel
spin), two to the para-species ($K_{\rm a}$=0, antiparallel spin).
A sixth line, also from para-H$_2$CO ($K_{\rm a}$ = 2) may have
been tentatively detected and may serve as a tracer of kinetic
temperature in future studies (M{\"u}hle et al. 2007). The rare
isotopologue HDCO was searched for in two transitions but remains
undetected. Our LVG calculations are based on collision rates of
Green (1991) with He that were scaled upwards by a factor of
1.37 to approximate collisions with H$_2$. Results are presented
in Fig.\,\ref{n113.c4}. Including 41 para- and 40 ortho-H$_2$CO
levels with molecular states up to 300\,cm$^{-1}$ above the ground
state, the results are consistent with optically thin emission.
The column density of ortho-H$_2$CO is about three times that of
para-H$_2$CO. This result is independent of the assumed kinetic 
temperature (20\,K$\leq$$T_{\rm kin}$$\leq$100\,K; Fig.\,\ref{n113.c4}). 
Since the lines show signal-to-noise ratios of order 10, 
1$\sigma$ errors are small and ortho-to-para abundance ratios below 
2.5 or above 3.5 can be excluded. The ratio agrees with that
expected in the case of formaldehyde formation in a warm ($T_{\rm
kin}$$\ga$40\,K) environment (e.g., Kahane et al. 1984; Dickens \&
Irvine 1999) and is not far off the value of 2.6, determined by
Heikkil{\"a} et al. (1999) for N\,159W. Densities vary between 
10$^{5}$ and 10$^{6}$\,cm$^{-3}$ but are, for a given kinetic 
temperature, within the uncertainties the same for ortho- and 
para-H$_2$CO.  This and similar linewidths (see Table~\ref{lin.para}) 
suggest that both formaldehyde species reside in the same volume.

While our LVG model results seem to agree with optically thin
emission, we do not have detections of rare isotopic species to
prove this. HDCO is weaker than the main species by a factor of
$\ga$30 (3$\sigma$). This is below the level found in the cool
environment of some low-mass protostellar cores (e.g., Roberts et
al. 2002; Parise et al. 2006) but does not provide significant
constraints, neither to the optical depths of the main species nor
to the cosmic D/H ratio (see e.g., Chin et al. 1996b; Heikkil{\"a}
et al. 1997; Gerin \& Roueff 1999 for details). With the lines
assumed to be optically thin, column densities become
$N$(para-H$_2$CO) $\sim$ 4$\times$10$^{12}$\,cm$^{-2}$ and
$N$(ortho-H$_2$CO) $\sim$ 1.2$\times$10$^{13}$\,cm$^{-2}$. These
column densities do not strongly depend on the assumed kinetic
temperature.

\subsubsection{CH$_3$OH}

Two $J$=2--1 and another two 3--2 lines of methanol (CH$_3$OH)
were detected (last panel of Fig.\,\ref{n113.5}). This is the
first time that the rotational multiplets of methanol have been
resolved in an extragalactic source (cf. Henkel et al. 1987;
Heikkil{\"a} et al. 1999). At kinetic temperatures of $T_{\rm
kin}$ = 50 and 100\,K, the measured profiles lead to $N$(CH$_3$OH)
$\sim$ 1.0$\times$10$^{13}$\,cm$^{-2}$ and $n_{\rm H_2}$ $\sim$
3.2$\times$10$^4$\,cm$^{-3}$ assuming optically thin emission. At
$T_{\rm kin}$ = 20\,K, the density is poorly determined and may
become $>$10$^5$\,cm$^{-3}$, while the column density is not
significantly changed.

\subsubsection{C$_3$H$_2$}

Both detected lines of cyclic C$_3$H$_2$ (hereafter c-C$_3$H$_2$)
belong to the ortho-species. Using an LVG code with collision rates
from Chandra \& Kegel (2000), we obtain densities of $n$(H$_2$) =
3$\times$10$^4$ and 9$\times$10$^4$\,cm$^{-2}$ for $T_{\rm kin}$ =
100 and 25\,K, respectively. The 1$\sigma$ uncertainty in the 
deconvolved line ratio, 0.39$\pm$0.08 (see Tables~\ref{lin.para} 
and \ref{lin.int}), which accounts for errors in the Gaussian fits, 
yields 1$\sigma$ deviations in density by a factor of 2. The 
column density becomes $N$(c-C$_3$H$_2$) $\sim$ 
3$\times$10$^{12}$\,cm$^{-2}$.

\subsection{Other molecular species}

\subsubsection{CN}

CN shows complex spectra. Each CN rotational state with $N$$>$0 is
split into a doublet by spin-rotation interaction. Because of the
spin of the nitrogen nucleus, each of these components is further
split into a triplet of hyperfine states. Calculated frequencies and
relative intensities are given by Skatrud et al. (1983).

CN was detected in a total of 10 lines, seven belonging to the 1--0
and three to the 2--1 transition. Relative intensities of the detected
$N$=1--0 features agree within 20\% with the LTE predictions. When
comparing its measured relative intensity with the corresponding LTE
value, the strongest line ($J$=3/2--1/2, $F$=5/2--3/2; weight:
$\sim$33\%) is weaker than expected by almost 20\%. For the two
lines with the second largest LTE weights ($\sim$12\%), such a
discrepancy is not apparent. We conclude that all CN $N$=1--0
features are not highly saturated, with the strongest feature having
an optical depth of $\tau$$\la$0.5.

The three $N$=2--1 observed transitions are also not providing 
evidence for high opacities. Here the line with intermediate intensity
($J$=3/2--1/2, $F$=5/2--3/2) is much weaker than expected. LTE
ratios are 26.7:16.7:6.1, which should be compared with observed
ratios of 31.5:10.6:7.4 (errors are of order 10--15\%).

In the optically thin limit, the CN excitation temperature can be
obtained from the ratio between the sum of the $N$=2--1 and 1--0
intensities using
$$
R_{21} = 4 \times\ {\rm e}^{-x} \times\ \frac{1-{\rm e}^{-2x}}{1-{\rm e}^{-x}} \times\
              \frac{({\rm e}^{2x}-1)^{-1} - ({\rm e}^{2y} - 1)^{-1}}{({\rm e}^{x}-1)^{-1} - ({\rm e}^{y} - 1)^{-1}},
$$
with $x$ = h$\nu_{10}$/k$T_{\rm ex}$, $y$ = h$\nu_{10}$/(k
$\times$ 2.73) and $\nu_{10}$ = 113.386\,GHz (see Wang et al.
2004). While the integrated intensity of the $N$=1--0 transition
is well determined (only two components representing together
$\sim$2.5\% of the total LTE intensity are not observed), the
three measured $N$=2--1 features only provide $\sim$50\% of the
total intensity under LTE conditions. Thus the total $N$=2--1
intensity in $R_{21}$ is less well determined and, in view of
potential deviations from LTE conditions, an error cannot be
derived. Ignoring these uncertainties and accounting for the LTE
intensities of the unobserved features, we obtain a line
intensity ratio of 1.20. Correcting for the different beam sizes
at 3 and 1.3\,mm yields $R_{21}$ = 0.69 and $T_{\rm ex}$ = 5.6\,K.
This value is small, indicating subthermal excitation (see e.g.,
Fuente et al. 1995). The column density becomes $N$(CN) $\sim$
3.5$\times$10$^{13}$\,cm$^{-2}$.

\subsubsection{Other molecules}

The remaining molecular species are seen in too few transitions and
do not provide enough information for a reliable estimate of density
and excitation. Assuming, however, that the lines are optically thin
(a realistic assumption, see Sects.\,4.2.2, 4.2.3, and 4.3.1) and
realizing that the column densities are almost independent of the
chosen kinetic temperature, column densities can still be estimated
in some cases. This involves an educated guess of the density.

We detected strong {\bf SO} emission in the 4$_3$--3$_2$ transition
(Fig.\,\ref{n113.5}). The column density (Table~\ref{tab.den}) was
calculated with an LVG code (70 levels up to 580\,K above the ground
state; collsion rates from Green 1994) for $T_{\rm kin}$ = 50\,K
and $n$(H$_2$) = 5$\times$10$^4$ and 5$\times$10$^5$\,cm$^{-3}$. The
resulting column density becomes
1.5$^{+1.5}_{-0.8}$$\times$10$^{13}$\,cm$^{-2}$.

Having observed two of the hyperfine components of the $N$=1--0
transition of {\bf C$_2$H}, we find that their line intensity ratio,
2.4$\pm$0.4, is consistent with 2.0, the expected value for
optically thin emission under LTE conditions. C$_2$H is present in
UV irradiated molecular clouds as well as in well shielded cores
(e.g., Pety et al. 2005; Beuther et al. 2008). Only having detected
the $N$=1--0 transition, and with no LVG code at hand, a reliable
column density could not be estimated (with the assumptions 
used by Wang et al (2004) for NGC\,4945, the source averaged 
5$\sigma$ upper limit would become 7$\times$10$^{14}$\,cm$^{-2}$).

{\bf N$_2$H$^+$} was detected in the $J$=1--0 line
(Fig.\,\ref{n113.3}). Individual hyperfine components are, unlike
those of HCN $J$=1--0, blended. For the column density
(Table~\ref{tab.den}) an LVG code was used with 31 levels up to
2100\,K above the ground state (collision rates from Sch{\"o}ier et
al. 2005). With the parameters also used for SO, a column density of
7$^{+1}_{-1}$$\times$10$^{11}$\,cm$^{-3}$ was calculated.

For {\bf SO$_2$}, detected in the 5$_{1,6}$--4$_{0,4}$ line, we used
RADEX, an LVG code also calculating the radiative transfer for a 
homogeneous, isothermal sphere with high velocity gradient (van der 
Tak et al. 2007). The column density becomes 
5$^{+1}_{-1}$$\times$10$^{12}$\,cm$^{-2}$.

{\bf HC$_3$N} was tentatively detected in the $J$=10--9 and 16--15
transitions (Fig.\,\ref{n113.4}). An LVG fit to the deconvolved
brightness temperatures (22 levels up to 92\,K above the ground
state; collision rates from Green \& Chapman 1978) yields for the
determined upper limits and $T_{\rm kin}$ = 50 and 100\,K densities
of $n$(H$_2$) $\sim$ 3$\times$10$^5$\,cm$^{-3}$ and
1.5$\times$10$^5$\,cm$^{-3}$, respectively. $N$(HC$_3$N) $\la$
6$\times$10$^{11}$\,cm$^{-2}$ (Table~\ref{tab.den}). A kinetic
temperature of 20\,K is too small to fit the two tentatively
detected lines.

Adopting the excitation conditions assumed by Mart\'{\i}n et al.
(2006) for NGC\,253, NO and HNCO source averaged 
5$\sigma$ upper limits to the column density would become 
6.5$\times$10$^{14}$\,cm$^{-2}$ and 2.6$\times$10$^{13}$\,cm$^{-2}$,
respectively.

\subsection{Densities and chemical implications}

The determination of spatial densities depends to a certain degree on
the assumed kinetic temperature. Nevertheless, Table~\ref{tab.den}
clearly shows that the molecular species analyzed in some detail
emit from regions that are characterized by three different density
regimes.  CO traces, not unexpectedly, the lowest density component,
C$_3$H$_2$ and CH$_3$OH originate from gas with intermediate
density. All the other molecules detected in more than one line
trace the high density component that can only occupy a small volume
because it is three orders of magnitude denser than the
$<$$n$(H$_2$)$>$ value derived for the inner 45\arcsec\ of the cloud
(Sect.\,4.2.1).

CO, the studied species with by far the highest column density, has
a natural advantage with respect to self-shielding in a hostile
interstellar environment like that of the LMC (see Sect.\,1).
Furthermore, a small dipole moment leads to a particularly low
``critical density''. Therefore collisional excitation to $T_{\rm
ex}$ values well above the level of the cosmic microwave background
is reached at comparatively low H$_2$ densities.

Cyclic C$_3$H$_2$ is a well known tracer of diffuse gas (e.g.,
Thaddeus et al. 1985; Cox et al. 1988) so that a lower density
than for other species with a similarly high dipole moment is
consistent with measurements of galactic clouds. However, such a
similarity is not self-evident. UV photons that are abundant in
the LMC (e.g., Israel et al. 1996) might destroy all c-C$_3$H$_2$
except in the densest regions. That this is not the case is an
important finding.

Quasi-thermal methanol (CH$_3$OH) has been observed to be sandwiched
between shocked regions, seen in vibrationally excited H$_2$, and
CS, tracing the dense ambient medium (Liechti \& Walmsley 1997; see
also Table~\ref{tab.den}). Fractional methanol abundances can vary
over several orders of magnitude (e.g., Kalenskii et al. 1997) and
are sometimes drastically enhanced by grain mantle evaporation so
that we may see emission arising preferentially from the outer, less
dense, highly UV irradiated edges of the molecular complex.

CS, HCN, HCO$^+$, and H$_2$CO trace the dense medium of N\,113,
but we should note that the CS, HCN, and HCO$^+$ $J$ = 1-0 lines
are slighty too strong for an optimal LVG fit. This suggests that
some of their emission is also arising from the component with
intermediate density. The $J$=1--0 line of para-H$_2$CO is not
part of the 3\,mm band and was therefore not observed. In galactic
clouds HCO$^+$ often arises from lower density gas than HCN
(e.g., Baan et al. 2008). Toward N\,113, we do not see 
this effect, which may imply that HCN and HCO$^+$ are efficiently 
destroyed by photodissociation in the lower density parts of the 
cloud. There is some evidence that H$_2$CO may be related to photon 
dominated regions, i.e. to the lower density edges of molecular 
clouds that are irradiated by UV-photons. Toward M\,82, another 
small galaxy with an intense UV field, Wei{\ss} et al. (2001) 
observed NH$_3$ and found that its emission must arise from 
relatively cool cloud cores that show a temperature similar to 
the overall dust radiation ($T_{\rm kin}$ $\sim$ 50\,K). H$_2$CO, 
however, was found to trace a much warmer gas component, with 
$T_{\rm kin}$ $\sim$ 200\,K and a surprisingly low density, 
$n$(H$_2$) $\sim$ 7$\times$10$^3$\,cm$^{-3}$ (M{\"u}hle et al. 
2007). In M\,82, H$_2$CO may be related to evaporated dust 
grain mantles (e.g., Ceccarelli et al. 2001), but apparently 
this scenario does not hold for N\,113.

Overall, the densities obtained for CO and the higher density
tracers are within the range estimated by Heikkil{\"a} et al.
(1999) for a number of other star forming clouds in the LMC. The
exception is HCO$^+$, which, in N\,113, traces H$_2$ densities
similar to those derived from HCN and CS, while Heikkil{\"a}
report lower H$_2$ densities. The high H$_2$ density obtained by
us is a surprise in view of the relatively large HCO$^+$ emission
region that was suggested by Wong et al. (2006; see our Sect. 4)
for N\.113 and by Heikkil{\"a} et al. (1999) for other clouds in
the LMC. It is possible, that only the HCO$^+$ $J$=1--0 line
originates from a relatively large volume, while the $J$=3--2
transition (see Fig.\,\ref{n113.4}) is solely tracing gas from the
densest regions.

\subsection{Column densities and chemical implications}

To derive the fractional abundances in Tables~\ref{tab.col.gal} 
and \ref{tab.col.ex} from the column densities given in Table~\ref{tab.den}, 
a reference is needed, which introduces an additional uncertainty. 
To minimize this error, we refer not to $N$(H$_2$) but provide
fractional abundances relative to $N$(CO). $N$(CO) has been analyzed 
in detail in Sect.\,4.2.1. What peculiarities can be expected? The
strongest deviations of abundances from galactic values may be 
related to nitrogen. Nitrogen is a predominantly ``secondary'' 
element. Being mainly synthesized on pre-existing carbon from a 
previous stellar generation, N underabundances are particularly 
large in regions having undergone little nuclear processing (e.g., 
Wheeler et al.  1989). B stars in the LMC show this effect. Observed 
are underabundances of C, N, and O by factors of $\sim$5, 10, and 
2 relative to the solar system (e.g., Hunter et al.  2007). 

The underabundance of N relative to C (i.e., CO) amounts to 
a factor of two. In view of the uncertainties mentioned above, this 
is close to the limit of what we can detect. Nevertheless, there are 
indications that this underabundance really plays a role. The two 
undetected species mentioned in Sect.\,3 (NO and HCNO) are 
both nitrogen bearing. Considering the spectral range between 150 
and 154\,GHz, H$_2$CO/HNCO and H$_2$CO/NO line ratios are $>$15 
and $>$20 versus $\sim$2.1 and $\sim$3.0 in NGC\,253 (see Mart\'{\i}n 
et al. 2006). For N\,113, we also find (Table~\ref{tab.den}) 
$N$(CS, SO, H$_2$CO, CH$_3$OH) $>$ $N$(HCN, HNC, N$_2$H$^+$, 
HC$_3$N), which is not typical for clouds outside the Magellanic
Clouds (Tables~\ref{tab.col.gal} and \ref{tab.col.ex}). 

Tables~\ref{tab.col.gal}--\ref{tab.col.ex} show that in N\,113 
HCN and HC$_3$N are clearly underabundant, relative to CO, with
respect to galactic as well as extragalactic molecular clouds
(N\,159 belongs like N\,113 to the LMC and shows a similar 
trend). The case for HNC is not as strong, but our HNC abundance 
is based on a single line only. It may seem surprising that 
the one species with two nitrogen atoms, N$_2$H$^+$, appears to be,
relative to CO, as abundant as in galactic cores and in the nuclear 
regions of large spiral galaxies (Tables~\ref{tab.col.gal} and 
\ref{tab.col.ex}). This may be explained by the fact that 
N$_2$H$^+$ is a tracer of dense cores. It is not as easily depleted 
in a very dense ($n$(H$_2$) $\sim$ 10$^6$\,cm$^{-3}$) medium as
CO and most other species (e.g., Fontani et al. 2006). In the LMC, 
where only cloud cores may be visible in a variety of molecules 
because of an otherwise insufficient shielding by dust grains, 
a low nitrogen abundance might thus be complemented by a source
averaged CO depletion of about the same amount, yielding an 
overall ``galactic'' N$_2$H$^+$/CO abundance ratio. We have 
to emphasize, however, that this result requires confirmation, 
because it is based on a single molecular N$_2$H$^+$ 
transition with undetermined optical depth.

To avoid as much as possible complications caused by uncommon
elemental abundances, relative column densities of isomers (e.g.,
HCN versus HNC) and of hydrogenated or protonated molecules (e.g.,
CN versus HCN, CO versus HCO$^+$ and H$_2$CO) are most suitable to
classify the molecular complex in terms of properties observed in
more metal-rich galactic molecular clouds. Notable is the 
high abundance of the cyanide radical CN relative to HCN and HNC. 
The former is to a large extent a photodissociation product of the 
latter two molecules. The high CN abundance is thus likely 
caused by the strong UV radiation field in the LMC (see, e.g., 
Fuente et al. 2006 for the related case of the starburst galaxy 
M\,82).

An $N$(HCN)/$N$(HNC) ratio larger than unity also favors the PDR 
scenario, which is further supported by 8$\mu$m emission attributed 
to polycyclic aromatic hydrocarbons (PAHs; see Wong et al. 2006) and 
by a lack of dust.  As already mentioned (Sect.\,4.1), the LMC 
requires for a given visual extinction about four times larger 
columns than the solar neighborhood (Bolatto et al. 2000). This 
allows UV radiation to penetrate deeper into the cloud. Associated 
with N\,113 are HD\,269219, a $\sim$30\,M$_{\odot}$ supergiant B 
star, HD\,269217, an emission line star, and a number of O9 to 
B0.5 stars either on the main sequence or slightly more evolved 
(Wilcots 1994; Oliveira et al. 2006). Table~\ref{tab.col.gal} 
presents fractional molecular abundances of a prototypical PDR, 
the Orion Bar.

Following Heikkil{\"a} et al. (1999), $N$(HC$_3$N)/$N$(CN) $<$ 0.1
is also indicative of a PDR. For N\,113 we find $\la$0.2 so that
in this case the result is inconclusive. The CO/HCO$^+$ and
$^{13}$CO/HCO$^+$ $J$=1--0 line intensity ratios, taken by
Heikkil{\"a} et al. (1999) as an inverse measure of star forming
and PDR activity (their Table~11), are with 13.8$\pm$0.4 and
1.7$\pm$0.1 a little higher than those of most other massive star
forming regions of the LMC. This suggests that PDR activity may be
less pronounced in N\,113. This and the prominent masers
(Sect.\,1) associated with the molecular complex may indicate that
massive star formation has been triggered more recently in N\,113
than in most other active cores of the LMC.

Aside from N\,113 (see also Chin et al. 1997; Wong et al. 2006)
N\,159 is the most thoroughly studied molecular complex of the LMC
(e.g., Heikkil{\"a} et al. 1999). Table~\ref{tab.col.ex} compares
column densities of several key molecules relative to $N$(CO).
Overall, nitrogen deficiencies may be similar in N\,159 and 
N\,113.

M\,82 is an irregular galaxy of similar size as the LMC, but it
hosts a starburst in its final stage. Accordingly, metallicities are
higher than in the LMC (e.g., Origlia et al. 2004). NGC\,253 and
NGC\,4945 are, on the other hand, large spirals with starburst
activity in their nuclear regions. Table~\ref{tab.col.ex} therefore
shows, from left to right, clouds with increasing metallicity.
Increasing column densities w.r.t. CO are found for HCN, HCO$^+$,
and HNC between Col.\,2 (N\,113) and Cols.\,5 and 6 (NGC\,253 and
NGC\,4945). For the nitrogen bearing species this is readily
explained by the nitrogen deficiency of clouds in the smaller
galaxies. It is not clear, however, why HCO$^+$ is following this
trend. There are only small differences in the column densities of
CS, H$_2$CO, and SO relative to those of CO. Considering CS/CO,
sulfur and oxygen are both synthesized in massive stars. H$_2$CO and
CO are both CO bearing. The $N$(SO)/$N$(CO) ratio is less trivial
because at least some of the carbon is, unlike sulfur, synthesized
in stars of intermediate mass. However, this may affect abundances
much less than the more notable underabundance of nitrogen.

\subsection{Isotope ratios}

Optically, it is difficult to discriminate between isotopes of a
given element, since their atomic lines are blended. However,
spectra from a given molecular species containing different isotopic
constituents, so-called ``isotopologues'', are well separated,
typically by a few percent of the rest frequency. This implies that
blending is no problem, while frequencies are still close enough to
be observed with the same technical equipment.

The hyperfine structure of the HCN $J$ = 1--0 line allows us to
determine the optical depth of each component and to obtain
accurate $^{12}$C/$^{13}$C, $^{14}$N/$^{15}$N and D/H ratios (for
the latter, see Chin et al. 1996b; Heikkil{\"a} et al. 1997). CO
provides $^{18}$O/$^{17}$O and even $^{16}$O/$^{18}$O ratios from
optically thin lines, while CS offers opportunities to study
sulfur isotopes.

Results are summarized in Table~\ref{tab.iso}. There are three major
findings: (1) {\it The ISM of the LMC is well mixed}, (2) {\it the
ratios are very different from the corresponding galactic values},
and (3) the carbon, nitrogen, and oxygen ratios demonstrate that
{\it the outer Galaxy is not providing an interstellar medium that
is intermediate between that of the solar neighborhood and the LMC}.

With respect to mixing, the $^{18}$O/$^{17}$O ratio is
particularly noteworthy because it has been determined in the most
direct way. Values of $\sim$1.6, much lower than in the Galaxy,
are measured in prominent star forming regions throughout the LMC
(Heikkil{\"a} et al. 1998). This does not only indicate that the
gas is quite homogeneous in its composition, but also excludes
effects of isotope selective fractionation, significant
differences in shielding against UV radiation, and C$^{17}$O line
intensities that are strongly influenced by non-LTE effects. Any
one of these might produce a notable scatter in the ratios, but
this is not observed.

The result that the ISM of the outer Galaxy does not provide a
direct connection between the solar neighborhood and the LMC, was
already addressed in the special case of $^{18}$O/$^{17}$O ratios
by Wouterloot et al. (2008) but deserves additional discussion.
Wouterloot et al. suggested on the basis of four ratios from the
outer Galaxy (galactocentric radii 16\,kpc $\leq$ $R_{\rm GC}$
$\leq$ 17\,kpc; solar radius $R_{\odot}$ = 8.5\,kpc) an
interpretation in terms of current models of galacto-chemical
evolution. Within the framework of ``biased-infall'' (e.g.,
Chiappini \& Matteucci 1999), the galactic disk is slowly formed
from inside out which is causing gradients in the abundances
across the disk. The average ratio appears to be $\sim$5 in the
outer disk but is only 1.6 in the LMC. In spite of the fact that
the outer disk $^{18}$O/$^{17}$O data are not numerous and
uncertainties are large, the difference is far too large not to be
significant. It is inconsistent with a pure metallicity dependence
that was proposed by Heikkil{\"a} et al. (1998). It also does not
indicate a lack of high mass stars as suggested by Heikkil{\"a} et
al. (1999), since massive stars are numerous in the LMC (e.g.,
Westerlund 1990). The high ratio of the outer Galaxy was
qualitatively interpreted in terms of a lack of $^{17}$O, that is
synthesized in largest quantities in stars of intermediate mass.
Their ejecta, reaching the ISM with a time delay, are less
dominant in the young stellar disk of the outer Galaxy than in the
older stellar body of the LMC (Wouterloot et al. 2008; see, e.g.,
Hodge 1989 for the star formation history of the LMC). This
interpretation is also consistent with the extremely high
$^{18}$O/$^{17}$O ratios ($>$10) obtained by Combes \& Wiklind
(1995) and Muller et al. (2006) toward the presumably young spiral
arms of galaxies seen at redshifts of $z$$\sim$0.7 and 0.9.

The carbon ratio offers another test of this scenario, also
indicating a well mixed interstellar medium in the LMC (see Sect.\,4.2.3). 
For $^{12}$C/$^{13}$C, the galactic data base is particularly large. 
$^{13}$C is like $^{17}$O a secondary nucleus and should be underabundant 
relative to $^{12}$C in the outer Galaxy, both with respect to the solar
neighborhood and the LMC. This is indeed the case (Table~\ref{tab.iso}).
$^{12}$C/$^{13}$C ratios are like $^{18}$O/$^{17}$O highest in the outer
Galaxy, smaller near the solar circle and smallest in the LMC. Also
the solar system $^{12}$C/$^{13}$C ratio is, like the $^{18}$O/$^{17}$O
ratio, higher than in the local interstellar medium. This is consistent
with an infusion of material from massive stars into the early solar
system and/or an enrichment of the local ISM by secondary nuclei during
the past 4.6$\times$10$^9$\,yr.

Less conclusive is the nitrogen ratio. Ratios from beyond the
Perseus arm ($R_{\rm GC}$$>$10\,kpc) have not yet been reported.
However, all galactic studies of elemental abundances or isotope
ratios extending to large radii show no signs for a change in radial
abundance gradients. If this also holds for $^{14}$N/$^{15}$N, then
we face again a situation with highest ratios in the outer Galaxy,
smaller ones near the solar system and smallest values in the LMC.
The interpretation would then be, like those for $^{18}$O/$^{17}$O
and $^{12}$C/$^{13}$C, that $^{15}$N is secondary with respect to
the main $^{14}$N species. This, however, is in conflict with a low
$^{14}$N/$^{15}$N ratio in the starburst galaxy NGC\,4945 which
suggests that $^{15}$N is a nucleus mainly synthesized in massive
stars due to rotationally induced mixing of protons into the
helium-burning shells of massive stars (Chin et al. 1999). In view
of this, there are two alternatives: (1) Either the weak HC$^{15}$N
profile presented by Chin et al. for NGC\,4945 is spurious and
$^{15}$N is predominantly synthesized in lower mass stars than
$^{14}$N or (2) the weak $^{14}$N/$^{15}$N gradient presented by
Wilson \& Rood (1994) is not significant and $^{15}$N is mainly a
product of massive stars. There are three arguments supporting the
latter view. These are (1) the low solar $^{14}$N/$^{15}$N ratio
(Table~\ref{tab.iso}) that may be understood in terms of the above
mentioned infusion of ejecta from massive stars, favoring $^{15}$N,
into the early solar system and/or enrichment of the local ISM 
by secondary nuclei (i.e., $^{14}$N) during the last 4.6$\times$10$^9$\,yr; 
(2) an extremely high $^{14}$N/$^{15}$N ratio ($>$600; Wilson \& Rood 
1994) in the ISM of the galactic center region that is dominated by 
products of CNO burning, mainly from stars of intermediate mass; 
(3) the likely low nitrogen isotope ratio in the young lensing 
galaxy of the PKS\,1830--211 system at redshift $z$$\sim$0.9 
(Muller et al. 2006). In case that the reported $^{14}$N/$^{15}$N 
galactic disk gradient is real, however, more complex interpretations 
will be required. Obviously, interstellar $^{14}$N/$^{15}$N ratios 
are not fully understood and more observational constraints are 
urgently needed.

Unlike $^{18}$O/$^{17}$O, $^{16}$O/$^{18}$O appears to be an
excellent tracer of metallicity. $^{16}$O and $^{18}$O are both
products of helium burning with metal poor stars apparently ejecting
little $^{18}$O. Highest $^{16}$O/$^{18}$O ratios are therefore
found in the LMC (Table~\ref{tab.iso}), lowest ratios in the
galactic center region. The entire range of observed interstellar
values covers almost an order of magnitude and would likely surpass
it, if C$^{18}$O could be detected in the Small Magellanic Cloud.

Chin et al. (1996a) reported a strong positive gradient with
galactocentric distance for $^{32}$S/$^{34}$S, which was a surprise
because both nuclei are synthesized by oxygen burning in massive
stars. The LMC ratio (Sect.\,4.2.2 and Table~\ref{tab.iso}) is 
substantially lower than those encountered in the local ISM and the 
solar system. Overall the situation seems to resemble that of the 
$^{14}$N/$^{15}$N ratio (a reported positive gradient in the galactic 
disk, a solar system ratio that is smaller than in the local ISM, 
and an LMC ratio that is smaller than ratios measured in the Galaxy) 
and thus provides a strong motivation for more detailed observational 
studies to constrain models of high mass stellar evolution.

\section{Conclusions}

Toward the prominent star-forming region N\,113 in the Large
Magellanic Cloud, we obtained with the SEST a map of the the
$\lambda$1.2\,mm dust continuum as well as spectral line data
covering 63 transitions from a total of 16 molecular species. These
include 50 detections, 7 tentative detections, and 6 undetected
transitions. The APEX and SEST telescopes also contribute 
molecular line maps.

(1) The total mass of the N\,113 molecular complex is estimated to
be a few 10$^5$\,M$_\odot$ with column densities of $N$(H$_2)$
$\sim$ 2--10$\times$10$^{22}$\,cm$^{-2}$ for the central 45\arcsec\
(11\,pc) and 24\arcsec (5.5\,pc), respectively. Assuming virial
equilibrium, we obtain for a 45\arcsec\ beam a $N$(CO)/$N$(H$_2$)
abundance ratio of about 4$\times$10$^{-5}$ and a conversion factor
of $X$ = $N$(H$_2$)/$I_{\rm CO 1-0}$ =
3.4$\times$10$^{20}$\,cm$^{-2}$ (K\,km\,s$^{-1}$)$^{-1}$.

(2) The virial theorem suggests an average density of $n$(H$_2$)
$\sim$ 500\,cm$^{-3}$ for the central 45\arcsec\ of the cloud. From
CO we obtain a density of $n$(H$_2$) $\sim$ 5000, for C$_3$H$_2$ and
CH$_3$OH the density becomes several 10$^4$\,cm$^{-3}$, while 
other molecular species (CS, HCN, HCO$^+$, and H$_2$CO)
mainly trace a gas density of several 10$^5$\,cm$^{-3}$. This 
indicates a high degree of clumping. Efficient shielding for 
most of the molecular species seems to occur in the densest regions 
only.

(3) Among the molecular species observed, only two remain
undetected, NO and HNCO. HC$_3$N is only tentatively detected. All
these molecules are nitrogen bearing. Chemically, the N\,113
molecular complex can be described by a photon dominated region in
an environment that lacks nitrogen. An ortho- to para-H$_2$CO column
density ratio of $\sim$3 indicates that at least formaldehyde was
formed in a warm ($T_{\rm kin}$ $\ga$ 40\,K) medium.

(4) Comparing isotope ratios with line intensity ratios we find that
CO is optically thick ($\tau$$\sim$10), while $^{13}$CO is optically
thin. The main lines of CS, HCN, and HCO$^+$ are only moderately
opaque ($\tau$$\sim$0.5).

(5) The interstellar medium of the LMC appears to be well mixed. 
Carbon, nitrogen, and oxygen isotope ratios demonstrate that the outer 
Galaxy does not provide a ``bridge'' between the interstellar medium 
of the solar neighborhood and that of the LMC. This is likely caused 
by the high age of the stellar population of the LMC relative to that 
of the outer Galaxy. Adopting this scenario, observed carbon and oxygen
isotope ratios are qualitatively understood, while nitrogen and
sulfur isotope ratios remain an enigma.

\acknowledgements We wish to thank S.~Leurini for the use of her
CH$_3$OH LVG code with collision rates provided by D.~Flower, 
M. Rubio for useful discussions, and an anonymous referee
for critically reading the manuscript. M.~W. acknowledges support
by the exchange program between the Chinese Academy of Sciences
and the Max-Plank-Gesellschaft, and partly by grants 10733030 
and 10621303 from NSFC and 2007CB815406 from MSTC. Y.-N.~C.
thanks for financial support by the National Science Council under
NSC91-2112-M-032-013. J.~B.~W. and M.~C. acknowledge the financial
support supplied by the Australian Government's Access to Major
Research Facilities Program (AMRFP) for travel to the SEST.

\clearpage

\begin{deluxetable}{cccccc}
\tablecolumns{5} \tablewidth{0pc}

\tablecaption{Spectroscopic parameters}

\tablehead{ \colhead{$\nu$} & 
        \colhead{Telescope} &
        \colhead{$\theta_{\rm b}$\tablenotemark{a)}} & 
        \colhead{$\theta_{\rm b}$\tablenotemark{a)}} & 
        \colhead{$T_{\rm sys}$\tablenotemark{b)}} &
        \colhead{$\eta_{\rm b}$\tablenotemark{c)}}
\\
\colhead{(GHz)} &            & \colhead{($''$)}        &
\colhead{(pc)} & \colhead{(K)} & \colhead{}}

\startdata

%         &            &               &                 &                 &              \\
 85--98   & SEST       &  61--53       &   15--13        &  180--270       & 0.77--0.74   \\
109--219  & SEST       &  47--24       &   11--5.8       &  250--400       & 0.70--0.48   \\
220--245  & SEST       &  23--21       &  5.6--5.1       &  550--1000      & 0.48--0.44   \\
265--268  & SEST       &  20--19       &  4.8--4.6       & $\sim$2000      & 0.41--0.40   \\
330--357  & SEST       &  16--14       &  3.9--3.4       & $\sim$3000      & 0.32--0.30   \\
345       &  APEX      &  20           &  4.8            &  $\sim$500      &  0.73   \\
%         &            &               &                 &                 &              \\
\enddata

\tablenotetext{a)}{Full Width to Half Power (FWHP) beam widths. To establish linear scales, $D$=50\,kpc was adopted.}
\tablenotetext{b)}{Single sideband system temperatures in units of main beam brightness temperature ($T_{\rm mb}$)}.  
\tablenotetext{c)}{SEST beam efficiencies were derived from measurements of Jupiter (L.~Knee, priv.  comm.). For the 
                   beam and forward hemisphere efficiency of APEX, see G{\"u}sten et al. (2006).}

\label{tel.param}
\end{deluxetable}

\clearpage

\begin{deluxetable}{ l r c r @{$\pm$} l r @{$\pm$} l r @{$\pm$} l r c }
\tablecolumns{11} \tablewidth{0pc}\tabletypesize{\scriptsize}

\tablecaption{Line parameters}

\tablehead{ \colhead{Transition}
     & \multicolumn{1}{c}{Frequency}
     & \colhead{Detection\tablenotemark{a)}}
     & \multicolumn{2}{c}{$\int$\,$T_{\rm mb}$\,d$v$\tablenotemark{b)}}
     & \multicolumn{2}{c}{$v_{\rm LSR}$\tablenotemark{c)}}
     & \multicolumn{2}{c}{$\Delta v_{1/2}$\tablenotemark{c)}}
     & \multicolumn{1}{c}{rms\tablenotemark{d)}}
     & \multicolumn{1}{c}{D$v$\tablenotemark{e)}} \\
     & \multicolumn{1}{c}{(MHz)}
     &
     & \multicolumn{2}{c}{(K\,km\,s$^{-1}$)}
     & \multicolumn{2}{c}{(km\,s$^{-1}$)}
     & \multicolumn{2}{c}{(km\,s$^{-1}$)}
     & \multicolumn{1}{c}{(K)}
     & \multicolumn{1}{c}{(km\,s$^{-1}$)} }

\startdata

c-C$_3$H$_2$ 2$_{1,2}$--1$_{0,1}$      & 85338.890   &+&0.20       & 0.04       & 235.92   & 0.18     & 2.83  &0.86       & 0.01       & 0.86\\
HC$^{15}$N 1--0                        & 86054.961   &? &0.03      & 0.01       & 234.76   & 0.27     & 3.96  &1.05       & 0.01       & 0.87\\
H$^{13}$CN 1--0                        & 86340.184   &+&0.08       & 0.01       & 236.76   & 0.52     & 7.41  &0.90       & 0.01       & 0.87\\
H$^{13}$CO$^+$ 1--0                    & 86754.294   &+&0.09       & 0.01       & 234.94   & 0.17     & 3.35  &0.46       & 0.01       & 0.89\\
C$_2$H 1--0 $J$=3/2--1/2 $F$=2--1      & 87316.925   &+&0.87       & 0.05       & 236.23   & 0.15     & 5.41  & 0.39      & 0.04       & 0.89\\
C$_2$H 1--0 $J$=3/2--1/2 $F$=1--0      & 87328.624   &+&0.36       & 0.05       & 235.71   & 0.35     & 4.76  & 0.74      & 0.04       & 0.89\\
HCN 1--0 $F$=1--1                      & 88630.416 & &0.81    & 0.01       &  235.04 & 0.04 & 4.80  & 0.04      & 0.01       & 0.85\\
HCN 1--0 $F$=2--1                      & 88631.847   &+& 1.26      & 0.01       & 235.04   & 0.04     & 4.80  &0.04       & 0.01       & 0.85\\
HCN 1--0 $F$=0--1                      & 88633.936 & &0.35    & 0.01       &  235.05 & 0.04 & 4.80  & 0.04      & 0.01       & 0.85\\
%4043 &HCN 1--0                        & 88631.847   &+&2.66       & 0.06       & 236.58   & 0.11     & 9.35  & 0.24      & 0.04       & 0.85\\
%4043 &HCN 1--0                        & 88631.847   & & 0.24      & 0.03       & 227.26   & 0.18     & 2.63  &0.29       & 0.04       & 0.85\\
%4043 &HCN 1--0                        & 88631.847   & &0.11       & 0.03       & 216.42   & 0.24     & 1.69  & 0.40      & 0.04       &0.85\\
HCO$^+$ 1--0                           & 89188.518   &+&3.56       & 0.09       & 235.12   & 0.07     & 5.67  &0.18       & 0.07       & 0.87\\
HNC 1--0                               & 90663.543   &+&0.94       & 0.04       & 235.23   & 0.11     & 5.14  & 0.23      & 0.02 & 0.99\\
HC$_3$N 10--9                          & 90978.993   &?&0.12       & 0.02       & 237.53   & 0.75     & 9.52  &1.38       & 0.01       & 0.96\\
$^{13}$CS 2--1                         & 92494.299   &+& 0.06      & 0.02       & 238.33   & 1.32     & 4.77  &2.25       & 0.01       & 0.98\\
N$_2$H$^+$ 1--0                        & 93173.404 &+&0.36   & 0.03       & 237.09 & 0.33 & 8.80  &0.69       & 0.02       & 0.97\\
C$^{34}$S 2--1                         & 96412.982   &+&0.15       & 0.01       & 235.21   & 0.19     & 4.39  &0.53       & 0.01       & 0.94\\
%2988 &CH$_3$OH 2--1                   & 96741.420   &+&0.17       & 0.02       & 235.07   & 0.28     & 3.89  &0.69       & 0.02       & 0.93\\
CH$_3$OH 2$_{-1}$--1$_{-1}$ E          & 96739.390   &+&0.10       & 0.02       & 235.90   & 0.31     & 3.58  &0.67       & 0.02       & 0.93\\
CH$_3$OH 2$_0$--1$_0$ A+               & 96741.420   &+&0.16       & 0.02       & 235.05   & 0.20     & 3.76  &0.44       & 0.02       & 0.93\\
CS 2--1                                & 97980.968   &+& 2.02      & 0.06       & 235.24   & 0.07     & 5.08  & 0.17      & 0.05       & 0.89\\
C$^{18}$O 1--0                         & 109782.160  &+&0.15       & 0.01       & 235.30   & 0.14     & 3.27  &0.35       & 0.01       & 0.94\\
$^{13}$CO 1--0                         & 110201.353  &+&  5.97     & 0.10       & 235.41   & 0.04     &  4.50 & 0.09      &  0.09      & 0.94\\
C$^{17}$O 1--0                         & 112358.988  &+& 0.09      & 0.01       & 235.01   & 0.42     &  7.25 & 0.75      &  0.01      & 0.92\\
CN 1--0, $J$=1/2--1/2 $F$=1/2--1/2     & 113144.122  &+& 0.27      & 0.03       & 234.83   & 0.30     &  5.47 & 0.74      &  0.02      & 0.91\\
CN 1--0, $J$=1/2--1/2 $F$=3/2--1/2     & 113170.502  &+& 0.38      & 0.03       & 235.10   & 0.17     &  5.03 & 0.43      &  0.02      & 0.91\\
%6441 &CN 1--0, $J$=1/2--1/2           & 113170.502  & & 0.29      & 0.06       & 283.00   & 0.10     &  16.39& 4.06      &  0.02      & 0.91\\
CN 1--0, $J$=1/2--1/2 $F$=3/2--3/2     & 113191.287  &+& 0.40      & 0.03       & 235.39   & 0.17     &  5.21 & 0.41      &  0.02      & 0.91\\
CN 1--0, $J$=3/2--1/2 $F$=3/2--1/2     & 113488.140  &+& 0.35      & 0.02       & 235.19   & 0.05     &  4.61 & 0.40      &  0.02      & 0.91\\
CN 1--0, $J$=3/2--1/2 $F$=5/2--3/2     & 113490.982  &+& 0.80      & 0.02       & 235.25   & 0.05     &  4.54 & 0.18      &  0.02      & 0.91\\
CN 1--0, $J$=3/2--1/2 $F$=1/2--1/2     & 113499.639  &+& 0.33      & 0.02       & 235.18   & 0.05     &  5.88 & 0.52      &  0.02      & 0.91\\
CN 1--0, $J$=3/2--1/2 $F$=3/2--3/2     & 113508.944  &+& 0.32      & 0.02       & 235.26   & 0.05     &  4.44 & 0.35      &  0.02      & 0.91\\
CO 1--0                                & 115271.204  &+&  49.21    & 0.31       & 235.25   & 0.02     &  5.79 & 0.04      &  0.25      & 0.90\\
HDCO 2$_{0,2}$--1$_{0,1}$              & 128812.860  &$-$ &\multicolumn{2}{c}{...}  &\multicolumn{2}{c}{...} & \multicolumn{2}{c}{...} & 0.02           & 0.88\\
HDCO 2$_{1,1}$--1$_{1,0}$              & 134284.910  &$-$ & \multicolumn{2}{c}{...} &\multicolumn{2}{c}{...} &\multicolumn{2}{c}{...}  & 0.01           & 0.93\\
SO$_2$ 5$_{1,5}$--4$_{0,4}$            & 135696.011  &+& 4.28      & 0.01       & 235.04   & 0.17     &  3.15 & 0.30      &  0.01      & 0.92\\
SO 4$_3$--3$_2$                        & 138178.648  &+& 1.15      & 0.04       & 235.01   & 0.07     &  4.56 & 0.18      &  0.04      & 0.93\\
$^{13}$CS 3--2                         & 138739.309  &+& 0.07      & 0.01       & 234.54   & 0.23     &  4.00 & 0.56      &  0.01      & 0.93\\
DCO$^+$ 2--1                           & 144077.321  &+& 0.17      & 0.01       & 235.37   & 0.20     &  5.17 & 0.42      &  0.01      & 0.89\\
H$_2$CO 2$_{1,2}$--1$_{1,1}$           & 140839.518  &+& 1.84      & 0.05       & 235.14   & 0.07     &  5.35 & 0.19      &  0.05      & 0.92\\
C$^{34}$S 3--2                         & 144617.147  &+& 0.23      & 0.02       & 234.82   & 0.11     &  3.29 & 0.27      &  0.02      & 0.89\\
DCN 2--1                               & 144828.000  &?& 0.06      & 0.02       & 234.45   & 1.14     &  7.07 & 1.49      & 0.01  & 0.89\\
CH$_3$OH 3$_{-1}$--2$_{-1}$ E          & 145097.470  &+& 0.31      & 0.02       & 234.92   & 0.12     &  4.23 & 0.44      &  0.02      & 0.89\\
CH$_3$OH 3$_0$--2$_0$ A+               & 145103.230  &+& 0.39      & 0.02       & 234.92   & 0.12     &  5.44 & 0.35      &  0.02      & 0.89\\
%2887 &CH$_3$OH 3--2                   & 145103.230  &+& 0.39      & 0.02       & 235.12   & 0.15     &  5.37 & 0.33      &  0.02      & 0.89\\
%2887    &CH$_3$OH 3--2                & 145103.230  &+& 0.30      & 0.02       & 246.64   & 0.13     &  4.00 & 0.38      &  0.02      & 0.89\\
%2482 &CH$_3$OH 3$_2$--2$_2$           & 145124.410  &$-$ &        &&&&&&& 1.44\\
HC$_3$N 16--15                         & 145560.946  &?& 0.09      & 0.02       & 235.82   & 0.43     &  3.71 & 0.88      & 0.01  & 0.89\\
H$_2$CO 2$_{0,2}$--1$_{0,1}$           & 145602.953  &+& 1.18      & 0.04       & 235.35   & 0.09     &  5.40 & 0.21      &  0.04      & 0.89\\
C$^{33}$S 3--2                         & 145755.620  &?& 0.06      & 0.01       & 238.38   & 0.53     & 4.65  & 1.20      & 0.01       & 0.86\\
CS 3--2                                & 146969.049  &+& 2.61      & 0.11       & 235.13   & 0.09     &  4.60 & 0.21      &  0.11      & 0.88\\
NO 2$\Pi_{1/2} J$=3/2--1/2 $F$=5/2--1/2& 150176.480  &$-$ &\multicolumn{2}{c}{...}&\multicolumn{2}{c}{...}&\multicolumn{2}{c}{...}&0.01& 0.86\\
H$_2$CO 2$_{1,1}$--1$_{1,0}$           & 150498.339  &+& 1.68      & 0.05       & 235.45   & 0.08     &  5.44 & 0.19      &  0.05      & 0.86\\
%6303 &C$_3$H$_2$ 4$_{0,4}$--3$_{1,3}$ \& 4$_{1,4}$--3$_{0,3}$ & 150835.000   &$-$ &           &&&&&&& 0.86\\
c-C$_3$H$_2$ 4$_{1,4}$--3$_{0,3}$      & 150851.910  &+& 0.15      & 0.01       & 235.25   & 0.19     & 3.63  & 0.32      &  0.01      & 0.86\\
HNCO 7$_{0,7}$--6$_{0,6}$              & 153865.092  &$-$ &\multicolumn{2}{c}{...}&\multicolumn{2}{c}{...}&\multicolumn{2}{c}{...}&0.02& 0.84\\
H$_2$CO 3$_{0,3}$--2$_{0,2}$           & 218222.188  &+& 0.62      & 0.05       & 233.24   & 0.19     &  4.68 & 0.41      &  0.02      & 0.96\\
H$_2$CO 3$_{2,2}$--2$_{2,1}$           & 218475.641  &$-$ &\multicolumn{2}{c}{...}&\multicolumn{2}{c}{...}&\multicolumn{2}{c}{...}&0.03& 0.92\\
%2704 &H$_2$CO 3$_{2,2}$--2$_{2,1}$           & 218475.641  &?& 0.08      & 0.04       & 236.19   & 0.70     &  2.04 & 1.04      &  0.08      & 0.92\\
H$_2$CO 3$_{2,1}$--2$_{2,0}$           & 218760.068  &$-$ &\multicolumn{2}{c}{...}&\multicolumn{2}{c}{...}&\multicolumn{2}{c}{...}&0.03& 0.96\\
$^{13}$CO 2--1                         & 220398.686  &+&  13.64    &  0.07      &  235.01  &  0.01    &   4.46&  0.03     &   0.09     & 0.91\\
H$_2$CO 3$_{1,2}$--2$_{1,1}$           & 225697.772  &+& 1.24      & 0.07       & 235.16   & 0.15     &  6.35 & 0.47      &  0.07      & 0.96\\
CN 2--1, $J$=3/2--1/2 $F$=5/2--3/2     & 226659.543  &+& 0.37      & 0.05       & 234.39   & 0.24     &  4.09 & 0.63      &  0.06      & 0.96\\
CN 2--1, $J$=3/2--1/2 $F$=3/2--1/2     & 226679.341  &+& 0.26      & 0.04       & 235.18   & 0.38     &  4.39 & 0.80      &  0.06      & 0.96\\
CN 2--1, $J$=5/2--3/2 $F$=7/2--5/2     & 226874.764  &+& 1.10      & 0.05       & 234.52   & 0.13     &  5.75 & 0.37      &  0.06      & 0.96\\
       CO 2--1                         & 230537.990  &+&  78.19    & 0.20       & 235.05  &  0.01     &  5.77&  0.02      &  0.24      & 0.87\\
C$^{34}$S 5--4                         & 241016.176  &?& 0.18      &0.04        & 232.30   & 0.47     &  4.20 & 0.76      & 0.02  & 0.93\\
CS 5--4                                & 244935.606  &+&  1.77     & 0.05       & 234.48   & 0.06     &  4.38 & 0.17      &  0.07      & 0.92\\
HCN 3--2                               & 265886.432  &+&  1.24     & 0.09       & 234.51   & 0.29     &   8.06&  0.66     &  0.09      & 0.79\\
HCO$^+$ 3--2                           & 267557.625  &+&  2.67     & 0.25       & 235.17   & 0.32     &  8.04 &  1.06     &  0.23      & 0.94\\
$^{13}$CO 3--2                         & 330587.957  &+&  17.73    & 0.65       & 234.80   & 0.08     &  4.78 & 0.23      &  0.94      & 0.91\\
CS 7--6                                & 342882.949  &?& 0.85      & 0.13       & 234.36   & 0.24     &  3.74 & 0.77      & 0.08  & 0.88\\
CO 3--2                                & 345795.975  &+&  79.76    & 0.53       & 235.10   & 0.02     &  6.16 & 0.05      &  0.75      & 0.87\\
HCO$^+$ 4--3                           & 356734.490  &+& 2.60      &0.22        &233.93    &0.22      & 5.47  &0.55       & 0.32       & 0.88\\

\enddata

\tablenotetext{a)}{`+': detection; `$-$': non-detection; `?':
tentative detection.}
\tablenotetext{b)}{Integrated from 220 to
240\,km\,s$^{-1}$ \ after subtracting a first order baseline or a
constant offset in $T_{\rm A}^{*}$. The errors were derived from
Gaussian fits and do not account for calibration and pointing
uncertainties.}
\tablenotetext{c)}{Obtained from single component
Gaussian fits. ${v_{\rm LSR}}$ =${v_{\rm HEL}}$ $-$4.5\,km\,s$^{-1}$
}
\tablenotetext{d)}{rms values for a $\sim $1\,km\,s$^{-1}$ \
channel width on a $T_{\rm mb}$ scale.} \tablenotetext{e)}{Channel
spacings after smoothing as shown in
Figs.\,\ref{n113.2}--\ref{n113.5}.}

\label{lin.para}
\end{deluxetable}

\clearpage

\begin{deluxetable}{l l l l l l l l }
\tablecolumns{8} \tablewidth{0pc}

\tablecaption{Half power source sizes of the 1.2\,mm continuum and
the mm-wave line emission (see Sect.\,3) observed with the SEST}

\tablehead{
\multicolumn{2}{l}{Transition} &
\multicolumn{1}{l}{$\theta_{\rm b}$\tablenotemark{a)}} &
\multicolumn{1}{l}{$\Delta\alpha$\tablenotemark{b)}} &
\multicolumn{1}{l}{$\Delta\delta$\tablenotemark{c)}} &
\multicolumn{1}{l}{$\theta_{\rm {s,\alpha}}$\tablenotemark{d)}} &
\multicolumn{1}{l}{$\theta_{\rm {s,\delta}}$\tablenotemark{d)}} &
\multicolumn{1}{l}{$\theta_{\rm s}$\tablenotemark{d)}}\\
\multicolumn{2}{l}{}& \multicolumn{6}{c}{($''$)}}

\startdata

250\,GHz              & Cont.    &24 &46 &46 &39 &39&                  39  \\
(1.2\,mm)             &          &   &   &   &   &  &                      \\
                      &          &   &   &   &   &  &                      \\
CO                    & 1--0     &45 &80 &70 &66 &54&                  60  \\
                      & 2--1     &23 &60 &60 &56 &56&                  56  \\
                      & 3--2     &15 &45 &40 &42 &37&                  40  \\
$^{13}$CO             & 1--0     &47 &65 &60 &45 &37&                  41  \\
                      & 2--1     &24 &40 &35 &32 &26&                  29  \\
CS                    & 3--2     &35 &42 &50 &23 &35&                  30  \\
                      & 5--4     &21 &40 &53 &34 &49&                  42  \\
HCO$^+$               & 1--0     &58 &60 &60 &15 &15&                  15  \\
HCN                   & 1--0     &58 &60 &54 &13 &---&                ---  \\
H$_2$CO               & 2--1     &37 &45 &54 &26 &39&                  33  \\

\enddata

\tablenotetext{a)}{$\theta_{\rm b}$, full width to half power (FWHP)
beam width at the transition observed}
\tablenotetext{b)}{$\Delta\alpha$, measured full half width in Right
Ascension} \tablenotetext{c)}{$\Delta\delta$, measured full half width in
Declination} \tablenotetext{d)}{$\Delta\alpha^2$ - $\theta_{\rm
b}^2$=$\theta_{\rm {s,\alpha}}^2$; $\Delta\delta^2$ - $\theta_{\rm
b}^2$=$\theta_{\rm {s,\delta}}^2$;
            $\theta_{\rm s}^2$=($\theta_{\rm {s,\alpha}}^2$ + $\theta_{\rm
            {s,\delta}}^2$)/2}

\label{cloudsize}

\end{deluxetable}

\clearpage

%\documentclass{aastex}
%\begin{document}

\begin{deluxetable}{l r c c r}
\tablecolumns{5} \tablewidth{0pc}\tabletypesize{\scriptsize}

\tablecaption{Line intensities}

\tablehead{ \colhead{Transition} & \multicolumn{1}{c}{$T_{\rm
mb}$~(K)} & \multicolumn{1}{c}{$\theta_{\rm s}$~($''$)} &
\multicolumn{1}{c}{$\eta_{\rm bf}$} & \multicolumn{1}{c}{$T^{
\prime}_{\rm mb}$~(K)}} %\hline

\startdata

CO   1--0                          &   8.853   &   60  &   0.640   &   13.833  \\
CO   2--1                          &   10.662  &   56  &   0.861   &   12.385  \\
CO   3--2                          &   12.171  &   40  &   0.877   &   13.884  \\
$^{13}$CO 1--0                     &   1.248   &   40  &   0.419   &   2.978   \\
$^{13}$CO 2--1                     &   2.244   &   40  &   0.743   &   3.021   \\
$^{13}$CO 3--2                     &   3.487   &   40  &   0.867   &   4.024   \\
C$^{18}$O 1--0                     &   0.044   &   40  &   0.417   &   0.105   \\
C$^{17}$O 1--0                     &   0.011   &   40  &   0.429   &   0.026   \\
    &       &       &       &       \\
CS   2--1                          &   0.415   &   40  &   0.363   &   1.141   \\
CS   3--2                          &   0.547   &   40  &   0.562   &   0.972   \\
CS   5--4                          &   0.281   &   40  &   0.781   &   0.359   \\
CS   7--6                          &   0.215   &   40  &   0.875   &   0.245   \\
C$^{34}$S 2--1                     &   0.032   &   40  &   0.356   &   0.091   \\
C$^{34}$S 3--2                     &   0.064   &   40  &   0.554   &   0.116   \\
C$^{34}$S 5--4                     &   0.041   &   40  &   0.775   &   0.053   \\
C$^{33}$S 3--2                     &   0.012   &   40  &   0.558   &   0.022   \\
$^{13}$CS 2--1                     &   0.012   &   40  &   0.337   &   0.037   \\
$^{13}$CS 3--2                     &   0.016   &   40  &   0.533   &   0.030   \\
    &       &       &       &       \\
C$_3$H$_2$ 2$_{1,2}$--1$_{0,1}$    &   0.058   &   40  &   0.302   &   0.193   \\
C$_3$H$_2$ 4$_{1,4}$--3$_{0,3}$    &   0.040   &   40  &   0.575   &   0.069   \\
H$_2$CO 2$_{1,2}$--1$_{1,1}$       &   0.333   &   40  &   0.541   &   0.616   \\
H$_2$CO 2$_{0,2}$--1$_{0,1}$       &   0.205   &   40  &   0.558   &   0.369   \\
H$_2$CO 2$_{1,1}$--1$_{1,0}$       &   0.289   &   40  &   0.574   &   0.504   \\
H$_2$CO 3$_{0,3}$--2$_{0,2}$       &   0.143   &   40  &   0.739   &   0.194   \\
H$_2$CO 3$_{1,2}$--2$_{1,1}$       &   0.183   &   40  &   0.752   &   0.244   \\
N$_2$H$^+$ 1--0                    &   0.039   &   40  &   0.340   &   0.114   \\
C$_2$H 1--0 3/2--1/2  F=2--1       &   0.152   &   40  &   0.312   &   0.486   \\
C$_2$H 1--0 3/2--1/2  F=1--0       &   0.070   &   40  &   0.312   &   0.226   \\
    &       &       &       &       \\
HNC  1--0                          &   0.172   &   40  &   0.328   &   0.525   \\
HCN   1--0                         &   0.158   &   40  &   0.319   &   0.495   \\
HCN   1--0                         &   0.260   &   40  &   0.319   &   0.815   \\
HCN   1--0                         &   0.068   &   40  &   0.319   &   0.213   \\
HCN   3--2                         &   0.160   &   40  &   0.815   &   0.196   \\
H$^{13}$CN  1--0                   &   0.009   &   40  &   0.307   &   0.029   \\
HC$^{15}$N  1--0                   &   0.007   &   40  &   0.305   &   0.022   \\
DCN    2--1                        &   0.010   &   40  &   0.555   &   0.018   \\
    &       &       &       &       \\
DCO$^+$ 2--1                       &   0.029   &   40  &   0.553   &   0.052   \\
H$^{13}$CO$^+$ 1--0                &   0.022   &   40  &   0.310   &   0.071   \\
HCO$^+$   1--0                     &   0.590   &   40  &   0.322   &   1.832   \\
HCO$^+$   3--2                     &   0.419   &   40  &   0.829   &   0.505   \\
HCO$^+$   4--3                     &   0.325   &   40  &   0.883   &   0.368   \\
HC$_3$N   10--9                    &   0.012   &   40  &   0.330   &   0.035   \\
HC$_3$N   16--15                   &   0.012   &   40  &   0.558   &   0.022   \\
SO$_2$ 5$_{1,5}$-4$_{0,4}$       &   0.013   &   40  &   0.523   &   0.025   \\
SO  4$_3$--3$_2$                   &   0.236   &   40  &   0.532   &   0.445   \\
CH$_3$OH 2$_0$--1$_0$ A$^+$        &   0.041   &   40  &   0.357   &   0.115   \\
CH$_3$OH 2$_0$--1$_0$ E$^+$        &   0.026   &   40  &   0.357   &   0.073   \\
CH$_3$OH 3$_0$--2$_0$ A$^+$        &   0.068   &   40  &   0.556   &   0.122   \\
CH$_3$OH 3$_0$--2$_0$ E$^+$        &   0.069   &   40  &   0.556   &   0.124   \\
    &       &       &       &       \\
CN 1--0 3/2--1/2 F=3/2--1/2        &   0.071   &   40  &   0.434   &   0.165   \\
CN 1--0 3/2--1/2 F=5/2--3/2        &   0.166   &   40  &   0.434   &   0.382   \\
CN 1--0 3/2--1/2 F=1/2--1/2        &   0.053   &   40  &   0.434   &   0.122   \\
CN 1--0 3/2--1/2 F=3/2--3/2        &   0.068   &   40  &   0.434   &   0.157   \\
CN 1--0 1/2--1/2 F=1/2--3/2        &   0.046   &   40  &   0.432   &   0.108   \\
CN 1--0 1/2--1/2 F=3/2--1/2        &   0.071   &   40  &   0.432   &   0.165   \\
CN 1--0 1/2--1/2 F=3/2--3/2        &   0.072   &   40  &   0.432   &   0.166   \\
CN 2--1 3/2--1/2 F=5/2--3/2        &   0.085   &   40  &   0.753   &   0.113   \\
CN 2--1 3/2--1/2 F=3/2--1/2        &   0.056   &   40  &   0.753   &   0.075   \\
CN 2--1 5/2--3/2 F=7/2--5/2        &   0.179   &   40  &   0.754   &   0.237   \\

\enddata

\label{lin.int}
\end{deluxetable}
%\end{document}

\clearpage

%\documentclass{aastex}
%\begin{document}

\begin{deluxetable}{lcccc}
\tablecolumns{5} \tablewidth{0pc}

\tablecaption{Source averaged logarithmic column densities and densities}

\tablehead{
\colhead{Molecule}           & \colhead{Column}        & \multicolumn{3}{c}{$n$(H$_2$) for} \\
                             & \colhead{density}       & \multicolumn{3}{c}{$T_{\rm kin}$\tablenotemark{a)}} \\
                             &               & \colhead{20\,K} & \colhead{50\,K} & \colhead{100\,K}             \\
                             & \colhead{(cm$^{-2}$)}    & \multicolumn{3}{c}{(cm$^{-3}$)}
                             }
%\hline

\startdata

CO                           &  17.8         & 3.7  & 3.7 & 3.7 \\
CN                           &  13.5         & ---  & --- & --- \\
CS                           &  13.4         & 6.0  & 5.7 & 5.3 \\
SO\tablenotemark{a)}         &  13.2         & ---  & --- & --- \\
SO$_2$\tablenotemark{a)}     &  12.7         & ---  & --- & --- \\
HCN                          &  12.8         & 6.0  & 5.6 & 5.3 \\
HCO$^+$                      &  12.6         & 5.8  & 5.6 & 5.3 \\
HNC                          &  12.4         & ---  & --- & --- \\
N$_2$H$^+$\tablenotemark{a)} &  11.8         & ---  & --- & --- \\
Para-H$_2$CO                 &  12.6         & 6.0  & 5.7 & 5.3 \\
Ortho-H$_2$CO                &  13.1         & 6.0  & 5.7 & 5.4 \\
c-C$_3$H$_2$                 &  12.5         & 5.0  & 4.7 & 4.5 \\
HC$_3$N\tablenotemark{b)}    &  11.8         & ---  & 5.5 & 5.2 \\
CH$_3$OH                     &  13.0         & 5.4  & 4.5 & 4.3 \\

\enddata

\tablenotetext{a)}{ Median for densities of 5$\times$10$^4$ and
5$\times$10$^5$\,cm$^{-3}$ and $T_{\rm kin}$ = 50\,K. See
Sect.\,4.3.2.}
\tablenotetext{b)}{ Based on two tentative detections. The given 
column density is thus a firm upper limit.}

\label{tab.den}

\end{deluxetable}
%\end{document}

\clearpage

%\documentclass{aastex}
%\begin{document}

\begin{deluxetable}{lccccc}
\tablecolumns{6} \tablewidth{0pc} \tablecaption{Source averaged
logarithmic column densities relative to CO toward N\,113 and
selected galactic targets\tablenotemark{a)}.}

\tablehead{
\colhead{Molecule}                     & \multicolumn{5}{c}{Column density (cm$^{-2}$)} \\
\cline{1-6}
                             &  \colhead{N\,113}  & \multicolumn{3}{c}{Orion}       & \colhead{TMC-1} \\
\cline{3-5}
                             &          & \colhead{Hot Core}         & \colhead{Ridge}     &  \colhead{Bar}  &
                             }
%                             &          & \colhead{Core}        &           &         \\
%                             & \multicolumn{5}{c}{(cm$^{-2}$)}

\startdata

CN                           &     --4.3 &      ---   &     --4.2 & --4.2 & --5.4 \\
CS                           &     --4.4 &     --4.3  &     --4.3 & --3.6 & --4.8 \\
SO                           &     --4.6 &     --3.3  &$\la$--4.7 & --4.0 & --5.1 \\
SO$_2$                       &     --5.1 &     --3.1  &  $<$--4.2 & --5.9 &  ---  \\
HCN                          &     --5.0 &     --2.9  &     --4.0 & --4.3 & --4.2 \\
HCO$^+$                      &     --5.2 &     --5.3  &     --4.3 & --4.5 & --4.3 \\
HNC                          &     --5.4 &     --5.1  &     --5.0 & --5.0 & --3.8 \\
N$_2$H$^+$                   &     --6.0 &      ---   &     ---   &  ---  & --5.8 \\
H$_2$CO                      &     --4.6 &     --4.5  &     ---   & --4.2 & --3.7 \\
c-C$_3$H$_2$                 &     --5.3 &      ---   &     ---   &  ---  & --4.0 \\
HC$_3$N                      &$\la$--6.0 &     --4.9  &     --5.6 &  ---  & --4.6 \\
CH$_3$OH                     &     --4.8 &     --3.2  &     ---   & --5.0 & --4.7 \\
%\hline

\enddata

\tablenotetext{a)}{Data for the Orion Hot Core and Ridge were
taken from Blake et al. (1987) and Comito et al. (2005).
Abundances for the Orion Bar refer to the (20\arcsec, --20\arcsec)
position of Jansen et al. (1995). TMC-1 values were adopted from
Leung et al. (1984), Madden et al. (1989), and Pratap et al.
(1997) for the cyanopolyne peak.}

\label{tab.col.gal}

\end{deluxetable}
%\end{document}

\clearpage

%\documentclass{aastex}
%\begin{document}

\begin{deluxetable}{lrrrrr}
\tablecolumns{6} \tablewidth{0pc}

\tablecaption{Source averaged logarithmic column densities relative
to CO toward N\,113 and other selected extragalactic
targets\tablenotemark{a)}.}

\tablehead{

\colhead{Molecule}                     & \multicolumn{5}{c}{Column density (cm$^{-2}$)}         \\
\cline{1-6}
                             &  \colhead{N\,113}  & \colhead{N\,159W}  &\colhead{M\,82}    &   \colhead{NGC}     & \colhead{NGC}       \\
                             &          &          &          &   \colhead{253}     &
                             \colhead{4945}
                             }
%                             & \multicolumn{5}{c}{(cm$^{-2}$)}            \\

\startdata

CN                           &    --4.3 &   --4.2  &   --4.3  &   --4.5   &     --4.3 \\
CS                           &    --4.4 &   --4.3  &   --3.8  &   --4.6   &     --4.1 \\
SO                           &    --4.6 &   --4.3  &$<$--4.4  &   --4.8   &     --4.7 \\
SO$_2$                       &    --5.1 &$<$--4.9  &   ---    &   ---     &      ---  \\
HCN                          &    --5.0 &   --4.6  &   --3.9  &   --3.9   &     --3.5 \\
HCO$^+$                      &    --5.2 &   --4.7  &   --4.2  &   --3.6   &$\ga$--5.4 \\
HNC                          &    --5.4 &   --5.1  &   --4.6  &   --4.2   &     --3.4 \\
N$_2$H$^+$                   &    --6.0 &   ---    &   --5.7  &   --5.8   &     --6.0 \\
H$_2$CO                      &    --4.6 &   --4.4  &   --3.4  &   --3.8   &     --4.5 \\
c-C$_3$H$_2$                 &    --5.3 &   --4.6  &   --4.6  &   --5.1   &     --4.1 \\
HC$_3$N                      &$\la$--6.0&$<$--5.2  &   --3.9  &   --3.8   &     --4.8 \\
CH$_3$OH                     &    --4.8 &   --4.4  &$<$--4.6  &   --3.8   &     --3.9 \\

\enddata

\tablenotetext{a)}{Assumed fractional CO abundance for N\,159W:
1.2$\times$10$^{-5}$; for M\,82, NGC\,253, NGC\,4945:
8$\times$10$^{-5}$. Data were taken from Heikkil{\"a} et al.
(1999), Wang et al. (2004) and Mart\'{\i}n et al. (2006).}

\label{tab.col.ex}
\end{deluxetable}

%\end{document}

\clearpage

%\documentclass{aastex}
%\begin{document}

\begin{deluxetable}{lccccccc}
\tablecolumns{6} \tablewidth{0pc}

\tablecaption{Isotope Ratios}

\tablehead{

\colhead{Isotope}           & \colhead{Solar}                     &
\colhead{Solar}             & \colhead{Outer}                     &
\colhead{LMC}               & \colhead{Proc.\tablenotemark{a)}}   &
\colhead{Ref.\tablenotemark{(b)}}                                 \\
\colhead{ratio}             & \colhead{system}                    &
\colhead{circle}            & \colhead{Galaxy}                    &
\colhead{(N113)}            &                                    \\}

\startdata

$^{12}$C/$^{13}$C   &   89   &  70    & 100           & 49$\pm$ 5     & 1 & 1,2              \\
$^{14}$N/$^{15}$N   &  270   & 400    &  --           & 91$\pm$21     & 2 & 1,3              \\
$^{16}$O/$^{18}$O   &  490   & 560    &  --           & 2000$\pm$250  & 3 & 1                \\
$^{18}$O/$^{17}$O   &  5.5   & 4.1    &   5           & 1.7$\pm$0.2   & 1 & 4                \\
$^{32}$S/$^{34}$S   &   22   &  28    &  --           & $\sim$15      & 4 & 5                \\
$^{34}$S/$^{33}$S   &    6   &   6    &  --           & $\geq$6       & 4 & 5                \\

\enddata

\label{tab.iso}

\tablenotetext{a)}{Primary processes of nucleosynthesis: (1)
Helium burning/CNO burning; (2) CNO burning; (3) Helium burning;
(4) Oxygen burning} \tablenotetext{b)}{References: (1) Wilson \&
Rood (1994); (2) Wouterloot \& Brand (1996); (3) Chin et al.
(1999); (4) Wouterloot et al. (2005, 2008); (5) Chin et al.
(1996a)}

\end{deluxetable}
%\end{document}

\clearpage

\clearpage

\clearpage

\clearpage

\clearpage

\clearpage

%\Online
%\appendix

%\section{Deconvolved line intensities}

\clearpage

%\section{Line maps}

\clearpage

%\section{Large Velocity Gradient (LVG) calculations}

\begin{thebibliography}{}
 \bibitem[2002]{1176}
  Aalto, S., Polatidis, A. G., H{\"u}ttemeister, S., \& Curran, S. J. 2002, A\&A, 381, 783
 \bibitem[2008]{1177}
  Baan, W. A., Henkel, C., Loenen, A. F., Baudry, A., \& Wiklind, T. 2008, A\&A, 477, 747
 \bibitem[1976]{1178}
  Bally, J., \& Langer, W. D. 1982, ApJ, 255, 143
 \bibitem[2008]{1180}
  Beuther, H., Semenov, D., Henning, T., \& Linz, H. 2008, A\&A, 675, L33
 \bibitem[1987]{1182}
  Blake, G. A., Sutton, E. C., Masson, C. R., \& Phillips, T. G. 1987, ApJ, 315, 621
 \bibitem[2000]{1184}
  Bolatto, A. D., Jackson, J. M., Israel, F., Zhang, X., \& Kim, S. 2000, ApJ, 545, 234
 \bibitem[1997]{1186}
  Brooks, K. J., \& Whiteoak, J. B. 1997, MNRAS, 291, 395
 \bibitem[1970]{1188}
  Castor, J. I. 1970, MNRAS, 149, 111
 \bibitem[2001]{1190}
  Ceccarelli, C., Loinard, L., Castets, A., Tielens, A. G. G. M., Caux, E., Lefloch, B., \& Vastel, C. 2001, A\&A, 372, 998
 \bibitem[2000]{1192}
  Chandra, S., \& Kegel, W. H. 2000, A\&AS, 142, 113
 \bibitem[1996]{1194}
  Chiappini, C., \& Matteucci, F. 1999, ApS\&S, 265, 425
 \bibitem[1996a]{1196}
  Chin, Y.-N., Henkel, C., Whiteoak, J. B., Langer, N., \& Churchwell, E. B. 1996a, A\&A, 305, 960
 \bibitem[1996b]{1198}
  Chin, Y.-N., Henkel, C., Millar, T. J., Whiteoak, J. B., \& Mauersberger, R. 1996b, A\&A, 312, L33
 \bibitem[1997]{1200}
  Chin, Y.-N., Henkel, C., Whiteoak, J. B., Millar, T. J., Hunt, M. R., \& Lemme, C. 1997, A\&A, 317, 548
 \bibitem[1998]{1202}
  Chin, Y.-N., Henkel, C., Millar, T. J., Whiteoak,~J.~B., \& Marx-Zimmer,~M. 1998, A\&A, 330, 901
 \bibitem[1999]{1204}
  Chin, Y.-N., Henkel, C., Langer, N., \& Mauersberger, R. 1999, ApJ, 512, L143
 \bibitem[1998]{1206}
  Cohen, R. S., Dame, T. M., Garay, G., Montani, J., Rubio, M., \& Thadeus, P. 1988, ApJ, 311, L95
 \bibitem[1995]{1208}
  Combes, F., \& Wiklind, T. 1995, A\&A, 303, L61
 \bibitem[2005]{1210}
  Comito, C., Schilke, P., Phillips, T. G., Lis, D. C., Motte, F., \& Mehringer, D. 2005, ApJS, 156, 127
 \bibitem[1988]{1212}
  Cox, P., G{\"u}sten, R., \& Henkel, C. 1988, A\&A, 206, 108
 \bibitem[1999]{1214}
  Dickens, J. E., \& Irvine, W. M. 1999, ApJ, 518, 733
 \bibitem[2001]{1216}
  Flower, D. R. 2001, J. Phys. B.: At. Mol. Opt. Phys., 34, 2731
 \bibitem[2006]{1217}
  Fontani, F., Caselli, P., Crapsi, A. et al. 2006, A\&A, 460, 709
 \bibitem[1982]{1218}
  Frerking, M. A., Langer, W. D., \& Wilson, R. W. 1982, ApJ, 262, 590
 \bibitem[1995]{1220}
  Fuente, A., Mart\'{\i}n-Pintado, J., \& Gaume, R. 1995, ApJ, 442, L33
 \bibitem[2006]{1221}
  Fuente, A., Garc\'{\i}a-Burillo, S., Gerin, M., et al. 2006, ApJ, 641, L105
 \bibitem[1999]{1222}
  Gerin, M., Roueff, E. 1999, in Highly Redshifted Radio Lines, ASP Conf. Ser. 156, eds. C. Carilli et al.
  San Francisco, p196
 \bibitem[1991a]{1225}
  Green, S. 1991, ApJS, 76, 979
 \bibitem[1991b]{1227}
  Green, S. 1994, ApJ, 434, 188
 \bibitem[1978]{1229}
  Green, S., \& Chapman, S. 1978, ApJS, 37, 169
 \bibitem[2006]{1231}
  G{\"u}sten, R., Nyman, L. A., Schilke, P., Menten, K. M., Cesarsky, C., \& Booth, R. 2006, A\&A 454, L13
 \bibitem[1997]{1233}
  Heikkil{\"a}, A., Johansson, L. E. B., \& Olofsson,~H. 1997, A\&A, 319, L21
 \bibitem[1998]{1235}
  Heikkil{\"a}, A., Johansson, L. E. B., \& Olofsson, H. 1998, A\&A, 332, 493
 \bibitem[1999]{1237}
  Heikkil{\"a}, A., Johansson, L. E. B., \& Olofsson, H. 1999, A\&A, 344, 817
 \bibitem[1993]{1239}
  Henkel, C., Mauersberger, R. 1993, A\&A 274, 730
 \bibitem[1987]{1241}
  Henkel, C., Jacq, T., Mauersberger, R., Menten, K. M., Steppe, H. 1987, A\&A, 188, L1
 \bibitem[1989]{1243}
  Hodge, P. 1989, ARA\&A, 27, 139
 \bibitem[2007]{1245}
  Hunter, I., et al. 2007, A\&A, 466, 277
 \bibitem[1996]{1247}
  Israel, F. P., Maloney, P. R., Geis, N., Herrmann, F., Madden, S. C., Poglitsch, A., \& Stacey, G. J.  1996, ApJ, 465, 738
 \bibitem[2003]{1249}
  Israel, F. P., et al. 2003, A\&A, 406, 817
 \bibitem[1995]{1251}
  Jansen, D. J., Spaans, M., Hogerheijde, M. R., \& van Dishoeck, E. F. 1995, A\&A, 303, 541
 \bibitem[1994]{1253}
  Johansson, L. E. B., Olofsson, H., Hjalmarson, A., Gredel, R., \& Black, J. H. 1994, A\&A, 291, 89
 \bibitem[1997]{1255}
  Kalenskii, S. V., Dzura, A. M., Booth, R. S., Winnberg, A., \& Alakoz, A. V. 1997, A\&A, 321, 311
 \bibitem[1984]{1257}
  Kahane, C., Lucas, R., Frerking, M. A., Langer, W. D., \& Encrenaz, P. 1984, A\&A, 137, 211
 \bibitem[2006]{1258}
  Klein, B., Philipp, S. D., Kr{\"a}mer, I., Kasemann, C., G{\"u}sten, R., Menten, K. M. 2006, A\&A 454, L29
 \bibitem[1984]{1259}
  Langer, W. D., Graedel, T. E., Frerking, M. A., \& Armentrout, P. B. 1984, ApJ, 277, 581
 \bibitem[1981]{1261}
  Larson, R. B. 1981, MNRAS, 194, 809
 \bibitem[2002]{1263}
  Lazendic, J. S., Whiteoak, J. B., Klamer, I., Harbison, P. D., \& Kuiper, T. B. H. 2002, MNRAS, 331, 969
 \bibitem[1984]{1265}
  Leung, C. M., Herbst, E., Huebner, W. F. 1984, ApJS, 56, 231
 \bibitem[1997]{1267}
  Liechti, S., \& Walmsley, C. M. 1997, A\&A, 321, 625
 \bibitem[1974]{1269}
  Lovas, F. J., \& Krupenie, P. H. 1974, J. Phys. Chem. Ref. Data, 3, 245
 \bibitem[1988]{1271}
  MacLaren, I., Richardson, K. M., \& Wolfendale, A. W. 1988, ApJ, 333, 821
 \bibitem[1997]{1273}
  Madden, S. C., Irvine, W. M., Swade, D. A., Matthews, H. E., \& Friberg, P. 1989, AJ, 97, 1403
 \bibitem[1993]{1275}
  Mangum, J. G. \& Wootten, A. 1993, ApJS, 89, 123
 \bibitem[2006]{1277}
  Mart\'{\i}n, S., Mauersberger, R., Mart\'{\i}n-Pintado, J., Henkel, C., \& Garc\'{\i}a-Burillo, S. 2006, ApJS, 164, 450
 \bibitem[1996a]{1280}
  Mauersberger,~R., Henkel,~C., Wielebinski, R., Wiklind, T., \& Reuter, H.-P. 1996a, A\&A, 305, 421
 \bibitem[1996b]{1282}
  Mauersberger,~R., Henkel,~C., Whiteoak,~J.~B., Chin,~Y.-N., \& Tieftrunk,~A.~R. 1996b, A\&A, 309, 705
 \bibitem[2004]{1284}
  Mauersberger, R., Ott, U., Henkel, C., Cernicharo, J. \& Gallino, R. 2004, A\&A 426, 219
 \bibitem[2005]{1288}
  Milam, S. N., Savage, C., Brewster, M. A., Ziurys, L. M., \& Wyckoff, S. 2005, ApJ, 634, 1126
 \bibitem[2006]{1290}
  Mizuno, N., Muller, E., Maeda, H., Kawamura, A., Minamidani, T., Onishsi, T., Mizuno, A., \& Fukui, Y.  2006, ApJ, 634, L107
 \bibitem[2007]{1292}
  M{\"u}hle, S., Seaquist, E. R., \& Henkel, C. 2007, ApJ, 671, 1579
 \bibitem[2006]{1294}
  Muller, S., Gu{\'e}lin, M., Dumke, M., Lucas, R., \& Combes, F. 2006, A\&A, 458, 417
 \bibitem[2006]{1295}
  Oliveira, J. M., van Loon, J. Th., Stanimirovi{\'c}, S., \& Zijlstra, A. A. 2006, MNRAS, 372, 1509
 \bibitem[2004]{1296}
  Origlia, L., Ranalli, P., Comastri, A., \& Maiolino, R. 2004, ApJ, 606, 862
 \bibitem[2006]{1298}
  Parise, B., Ceccarelli, C., Tielens, A. G. G. M., Castets, A., Caux, E., Lefloch, B., \& Maret, S. 2006, A\&A, 453, 949
 \bibitem[1981]{1300}
  Penzias, A. A. 1981, ApJ, 249, 518
 \bibitem[2005]{1302}
  Pety, J., Teyssier, D., Foss{\'e}, D., Roueff, E., Abergel, A., Habart, E., \& Cernicharo, J. 2005, A\&A, 435, 885
 \bibitem[1997]{1304}
  Pratap, P., Dickens, J. E., Snell, R. L., Miralles, M. P., Bergin, E. A., Irwin, W. M., \& Schloerb, F. P. 1997, ApJ, 486, 862
 \bibitem[2006]{1305}
  Risacher, C. et al. 2006, A\&A, 454, L17
 \bibitem[2002]{1306}
  Roberts, H., Fuller, G.A., Millar, T. J., Hatchell, J., \& Buckle, J. V. 2002, A\&A, 381, 1026
 \bibitem[1991]{1308}
  Rubio, M., Garay, G., Montani, J., \& Thaddeus, P. 1991, ApJ, 368, 173
 \bibitem[1993]{1310}
  Rubio, M., Lequeux, J., \& Boulanger, F. 1993, A\&A, 271, 9
 \bibitem[1992]{1316}
  Schilke, P., Walmsley, C. M., Pineau de For{\^e}ts, G., Roueff, E., Flower, D. R., \& Guilloteau, S. 1992, A\&A, 256, 595
 \bibitem[2005]{1312}
  Sch{\"o}ier, F. L., van der Tak, F. F. S., van Dishoeck, E. F., \& Black, J. H. 2005, A\&A, 432, 369
 \bibitem[1974]{1314}
  Scoville, N. Z., \& Solomon, P. M. 1974, ApJ, 187, L67
 \bibitem[1987]{1318}
  Skatrud, D. D., De Lucia, F. C., Blake, G. A., \& Sastry, K. V. L. N. 1983, J. Mol. Spec., 99, 35
 \bibitem[1960]{1320}
  Sobolev, V. V. 1960, in Moving Envelopes of Stars (Cambridge, Harvard University Press)
 \bibitem[1985]{1322}
  Thaddeus, P., Vritilek, J. M., \& Gottlieb, C. A. 1985, ApJ, 299, 63
 \bibitem[1992]{1324}
  Turner, B. E., Chan, K.-W., Green, S., \& Lubowich, D. A. 1992, ApJ, 399, 114
 \bibitem[2007]{1326}
  van der Tak, F. F. S., Black, J. H., Sch{\"o}ier, F. L., Jansen, D. J., \& van Dishoeck, E. F. 2007, A\&A, 468, 627
 \bibitem[2004]{1328}
  Wang, M., Henkel, C., Chin, Y.-N., Whiteoak, J. B., Hunt Cunningham, M., Mauersberger, R., \& Muders, D. 2004, A\&A, 422, 883
 \bibitem[1976]{1330}
  Watson, W. D., Anicich, V. G., \& Huntress, W. T. 1976, ApJ, 205, L165
 \bibitem[2001]{1332}
  Wei{\ss}, A., Neininger, N., Henkel, C., Stutzki, J., \& Klein, U. 2001, A\&A, 554, 143
 \bibitem[1990]{1334}
  Westerlund, B. E. 1990, A\&AR, 2, 29
 \bibitem[1989]{1336}
  Wheeler, J. C., Sneden, C., \& Truran, J. W. 1989, ARA\&A, 27, 279
 \bibitem[1986]{1338}
  Whiteoak, J. B., \& Gardner, F. F. 1986, MNRAS, 222, 513
 \bibitem[1994]{1339}
  Wilcots, E. M. 1994, AJ, 108, 1674
 \bibitem[1994]{1340}
  Wilson, T. L., \& Rood, R. 1994, ARA\&A, 32, 191
 \bibitem[2006]{1342}
  Wong, T., Whiteoak, J. B., Ott, J., Chin, Y.-N., \& Cunningham, M. R. 2006, ApJ, 649, 224
 \bibitem[1996]{1345}
  Wouterloot, J. G. A., \& Brand, J. 1996, A\&AS, 119, 439
 \bibitem[1986]{1347}
  Wouterloot, J. G. A., \& Walmsley, C. M. 1986, A\&A, 168, 237
 \bibitem[2005]{1349}
  Wouterloot, J. G. A., Brand, J., \& Henkel, C. 2005, A\&A, 430, 549
 \bibitem[2008]{1351}
  Wouterloot, J. G. A., Henkel, C., Brand, J., \& Davis, G. R. 2008, A\&A, 487, 237
 \bibitem[2001]{1353}
  Yamaguchi, R., Mizuno, N., Onishi, T. Mizuno, A., \& Fukui, Y.  2001, PASJ, 53, 985
\end{thebibliography}
\end{document}